\documentclass[twocolumn]{aastex7}
\usepackage{graphicx}
\usepackage{subcaption} 
\usepackage{amsmath} 
\usepackage{bm}       
\usepackage{hyperref} 
\usepackage{booktabs}   
\usepackage{contour} 
\usepackage{adjustbox} 
\usepackage{natbib}     

\begin{document}

\title{A Method for Gamma-Ray Energy Spectrum Inversion and Correction}

\correspondingauthor{Xin-Qiao Li}
\email{lixq@ihep.ac.cn}

\author[0009-0005-5459-3600]{Zhi-Qiang Ding}
\affil{State Key Laboratory of Particle Astrophysics, Institute of High Energy Physics, Chinese Academy of Sciences, Beijing 100049, China}
\affil{University of Chinese Academy of Sciences, Chinese Academy of Sciences, Beijing 100049, China}
\email{}  

\author[orcid=]{Xin-Qiao Li*}
\affil{State Key Laboratory of Particle Astrophysics, Institute of High Energy Physics, Chinese Academy of Sciences, Beijing 100049, China}
\email[show]{} 

\author[orcid=0000-0003-4311-5804]{Da-Li Zhang}
\affil{State Key Laboratory of Particle Astrophysics, Institute of High Energy Physics, Chinese Academy of Sciences, Beijing 100049, China}
\email{} 

\author[orcid=]{Zheng-Hua An}
\affil{State Key Laboratory of Particle Astrophysics, Institute of High Energy Physics, Chinese Academy of Sciences, Beijing 100049, China}
\email{}

\author[orcid=]{Zhen-Xia Zhang}
\affil{National Institute of Natural Hazards, Ministry of Emergency Management of the People’s Republic of China, Beijing 100085, China}
\email{} 

\author[orcid=]{Roberto Battiston}
\affil{University of Trento and INFN, Trento 38123, Italy}
\email{} 

\author[orcid=]{Roberto Iuppa}
\affil{University of Trento and INFN, Trento 38123, Italy}
\email{} 

\author[orcid=]{Zhuo Li} 
\affil{Department of Astronomy, School of Physics, Peking University, Beijing 100871, China}
\affil{Kavli Institute for Astronomy and Astrophysics, Peking University, Beijing 100871, China}
\email{}

\author[0000-0001-5348-7033]{Yan-Qiu Zhang}
\affil{State Key Laboratory of Particle Astrophysics, Institute of High Energy Physics, Chinese Academy of Sciences, Beijing 100049, China}
\affil{University of Chinese Academy of Sciences, Chinese Academy of Sciences, Beijing 100049, China}
\email{}

\author[orcid=]{Yan Huang} 
\affiliation{School of Physics and Optoelectronics Engineering, Anhui University, Hefei 230601, China}
\email{}

\author[0009-0001-7226-2355]{Chao Zheng}
\affil{State Key Laboratory of Particle Astrophysics, Institute of High Energy Physics, Chinese Academy of Sciences, Beijing 100049, China}
\affil{University of Chinese Academy of Sciences, Chinese Academy of Sciences, Beijing 100049, China}
\email{}

\author[]{Yan-Bing Xu}
\affil{State Key Laboratory of Particle Astrophysics, Institute of High Energy Physics, Chinese Academy of Sciences, Beijing 100049, China}
\email{}

\author[]{Xiao-Yun Zhao}
\affil{State Key Laboratory of Particle Astrophysics, Institute of High Energy Physics, Chinese Academy of Sciences, Beijing 100049, China}
\email{}

\author[]{Lu Wang}
\affil{State Key Laboratory of Particle Astrophysics, Institute of High Energy Physics, Chinese Academy of Sciences, Beijing 100049, China}
\affil{University of Chinese Academy of Sciences, Chinese Academy of Sciences, Beijing 100049, China}
\email{}

\author[]{Ping Wang}
\affil{State Key Laboratory of Particle Astrophysics, Institute of High Energy Physics, Chinese Academy of Sciences, Beijing 100049, China}
\email{}

\author[]{Hong Lu}
\affil{State Key Laboratory of Particle Astrophysics, Institute of High Energy Physics, Chinese Academy of Sciences, Beijing 100049, China}
\email{}


\begin{abstract}
Accurate spectral analysis of high-energy astrophysical sources often relies on comparing observed data to incident spectral models convolved with the instrument response. However, for Gamma-Ray Bursts and other high-energy transient events observed at high count rates, significant distortions (e.g., pile-up, dead time, and large signal trailing) are introduced, complicating this analysis. We present a method framework to address the model dependence problem, especially to solve the problem of energy spectrum distortion caused by instrument signal pile-up due to high counting rate and high-rate effects, applicable to X-ray, gamma-ray, and particle detectors. Our approach combines physics-based Monte Carlo (MC) simulations with a model-independent spectral inversion technique. The MC simulations quantify instrumental effects and enable correction of the distorted spectrum. Subsequently, the inversion step reconstructs the incident spectrum using an inverse response matrix approach, conceptually equivalent to deconvolving the detector response. The inversion employs a Convolutional Neural Network, selected for its numerical stability and effective handling of complex detector responses. Validation using simulations across diverse input spectra demonstrates high fidelity. Specifically, for 27 different parameter sets of the brightest gamma-ray bursts, goodness-of-fit tests confirm the reconstructed spectra are in excellent statistical agreement with the input spectra, and residuals are typically within $\pm 2\sigma$. This method enables precise analysis of intense transients and other high-flux events, overcoming limitations imposed by instrumental effects in traditional analyses.

\end{abstract}

\section{Introduce} \label{sec:1}

Gamma-Ray Bursts (GRBs) rank among the most energetic astrophysical phenomena identified to date \citep{fishman1995gamma}, manifesting as extraordinary releases of energy from cataclysmic cosmic events \citep{piran2004physics, kumar2015physics}. The immense energy radiated by a GRB in a brief period of seconds can exceed the total energy output of the Sun over its entire lifetime by orders of magnitude \citep{Zhang_2018}. Consequently, the observation and study of GRBs are widely recognized as playing a crucial role across numerous frontiers in astronomy and astrophysics \citep{woosley2006supernova}. Basic physics issues related to GRBs, such as the equation of state of ultra-dense matter, stellar evolution through the collapse of massive stars and the production of kilonovae and supernovae, jet formation and dissipation, the generation and propagation of gravitational waves \citep{abdalla2022cosmology}, and the pursuit of key cosmological parameters \citep{demianski2017cosmology}. In recent years, GRBs have re-emerged as critical cosmic messengers in multi-messenger astrophysics, spurred by their potential association with kilonova transients and gravitational wave events \citep{abbott2017gravitational, troja2022nearby, dichiara2023luminous}. Precise measurements of GRB energy spectra are crucial for addressing these fundamental astrophysical questions. However, the inherent model dependence of spectral analysis, coupled with a combination of instrumental effects---namely signal pile-up, dead time, and large signal trailing---poses considerable challenges to the identification and investigation of new spectral morphologies, especially for bright GRBs. Hereafter, we use the term ``instrumental effects'' to refer to these combined phenomena. 

First, standard spectral analysis methods, including those implemented in widely-used packages like XSPEC \citep{1996ASPC..101...17A}, typically employ forward-fitting techniques. These methods convolve an assumed physical model with the energy response matrix and compare the result to the observed counts \citep{fishman1995gamma, 2002ApJ...581.1236Z}. Although useful for model testing, this approach cannot provide model-independent spectral reconstructions. The spectral inversion problem is inherently ill-posed \citep{yuan2009ill, bakushinsky2010iterative}, rendering direct solutions vulnerable to noise amplification and consequently producing substantial uncertainties in the derived spectral parameters. Second, for bright transients like GRBs, the extreme photon fluxes cause significant deviations from standard calibration procedures that typically use the energy response matrix and average dead time corrections \citep{2014ExA....38..433Z, 2020JHEAp..26...58X}. At peak flux levels \citep{bhat2014fermi}, \textbf{these instrumental effects} are highly nonlinear and dynamically coupled, introducing substantial spectral distortions \citep{zahn2019pile, chaplin2013analytical}. These complex instrumental responses create significant discrepancies between observed and incident spectra.

The correction of instrumental effects is a well-established challenge in high-energy astrophysics. In X-ray astronomy, a rich history of development exists for mitigating detector artifacts, where even a well-characterized effect like photon pile-up can be ``challenging to model analytically'' \citep{2025AAS...24513306Y}. The traditional paradigm has progressed from early analytical formalisms \citep{1997ChNew...5...14K, 1998SPIE.3444...30B, 1999A&AS..135..371B} to sophisticated statistical models integrated into spectral fitting software \citep{2000HEAD....5.2706D, 2004HEAD....8.1635Y, 2005ASPC..347..478M}, with a recent shift toward data-driven, neural network-based emulators \citep{2025AAS...24513306Y}. This problem, however, is compounded in high-energy gamma-ray detectors, where pile-up is coupled with other significant non-linearities, such as detector dead time and large signal trailing.

Current space-borne missions like \textit{Fermi}/GBM and GECAM \citep{li2022technology, an2022design} illustrate the limitations of conventional techniques in this regime. These instruments employ state-of-the-art, model-based corrections. For instance, \textit{Fermi}/GBM utilizes a model that predicts recorded counts and spectra from a constant-intensity Poisson process affected by pile-up, validated via radioactive source experiments \citep{2013NIMPA.717...21C, 2014ExA....38..331B}. Similarly, the GECAM mission constructed a high-precision response matrix (RMF) for its detectors (energy range 6\,keV to 6\,MeV) through extensive Monte Carlo simulations and comprehensive ground and in-flight calibrations \citep{2023NIMPA105668586Z, 2024NIMPA105969009Z}. Despite their sophistication, these approaches inherently rely on simplifying assumptions---such as channel independence and steady-state count rates---that are often violated during dynamic, high-rate events like Gamma-Ray Bursts. Crucially, they fail to decouple the intertwined instrumental effects inherent in the full detector chain response, from incident photon energy deposition \citep{xiao2016crosstalk} to signal shaping and data acquisition (DAQ). Furthermore, traditional matrix deconvolution algorithms for spectral inversion, such as Tikhonov regularization, are prone to noise amplification when applied to such ill-posed inverse problems, leading to substantial uncertainties in the recovered spectra \citep{hager2015alternating, calvetti2025distributed}. Our work therefore contributes to the emerging data-driven trend by introducing a comprehensive framework that uses machine learning to correct for this full, coupled suite of instrumental effects in the gamma-ray regime.

The High Energy Particle Package (HEPP) is one of the main payloads of the China Seismo-Electromagnetic Satellite (CSES). CSES aims to monitor space electromagnetic fields, ionospheric plasma, high-energy charged particles, and other features of the global space environment \citep{2019EGUGA..21.3809C}. Developed by the Institute of High Energy Physics (IHEP), Chinese Academy of Sciences, HEPP-H - the high-energy detector of HEPP \citep{2019RDTM....3...22L} - detects electrons (1.5-50 MeV) and protons (15-200 MeV) across a broad energy range. Crucially, HEPP-H also incorporates a rate meter sampling capability, characterized by a dead time of 0.1~$\mu$s, which allows for dead-time corrections using its data. In addition to primary charged particles, HEPP can also detect secondary particles generated by the interaction between gamma rays and satellite materials \citep{2023ApJ...946L..29B}. However, in our analysis of GRB observations acquired by the HEPP-H instrument, we identified the need to implement corrections for instrumental effects related to the detector's signal acquisition process.

Therefore, this paper presents two approaches. First, we propose to employ MC simulations to derive a correction function and achieve convergence in spectral fits through an iterative fitting cycle incorporating this correction function. We acknowledge that the derived correction function might exhibit some dependence on the initial spectral assumptions used in the MC simulation. To validate the reliability of this iterative correction, we further introduce a novel inversion method. This method aims to construct an ``inverse response matrix'', which is independent of the incident energy spectrum, over a broad energy range, enabling direct inversion of the detected spectrum to recover the incident spectrum. We validate the proposed inversion method extensively using simulations and observational data. Subsequently, we utilize this validated inversion method to systematically evaluate the systematic error in the MC-derived correction function.

The core ideas supporting the above calibration and verification framework mainly lie in the following three aspects:

(1) \textbf{Monte Carlo model:} We have developed a MC methodology \citep{brooks1998markov}. This approach meticulously simulates, step-by-step, the physical processes from the incident photon's time-energy distribution through energy deposition, signal waveform superposition (Landau function), and field programmable gate array (FPGA) state machine dead time logic. This implementation enables the accurate quantification of \textbf{the non-linear instrumental effects.}

(2)  \textbf{Inverse Energy Response Matrix:} This method reconstructs the incident particle spectrum by reformulating the relationship between incident and detected spectra. Instead of modeling the detected spectrum as a response to the incident spectrum, we mathematically invert this relationship, effectively interchanging their conventional roles to express the incident spectrum as a function of the detected spectrum. This allows for the construction of an inverse energy response matrix conceptually designed to map the measured spectrum back to the incident one. This matrix formalism is developed with the goal of minimizing dependence on specific spectral shapes assumed during its derivation. The reconstructed incident spectrum is then obtained by applying the effective inverse response operation to the measured detector spectrum. We implement this inversion using a Convolutional Neural Network (CNN; \cite{o2015introduction, wu2017introduction}), trained to robustly learn the transformation from the detected spectrum to the reconstructed incident spectrum.

(3) \textbf{Construction of energy spectrum distortion correction function and systematic error analysis:} This work culminates in the development of a spectral distortion correction function, rigorously validated through 27 cross-validation experiments spanning the spectral parameter space of the brightest GRB observed to date \citep{2023arXiv230301203A}. The results from rigorous goodness-of-fit tests show that the corrected and true spectra are statistically consistent across the vast majority of test cases, effectively ensuring the general applicability of the framework to diverse spectral morphologies of the GRBs. Furthermore, this comprehensive validation provides a quantified upper bound on the systematic error of the correction framework.

\section{Data Acquisition Correction} \label{sec:2}

\begin{figure}[h] 
  \centering 
  \includegraphics[width=0.47\textwidth]{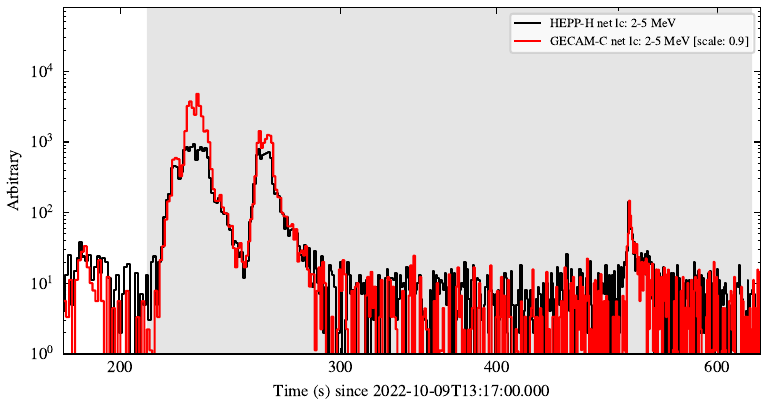}
  \caption{Light curves from HEPP-H (black) and GECAM-C (red) in the 2--5~MeV energy range. The discrepancies between the two curves highlight the instrumental effects present in the HEPP-H data. The deviations observed in the HEPP-H light curve are attributed to comprehensive instrumental effects, including particle incidence properties, detector energy response, large signal trailing, pile-up, and data acquisition dead time. Such instrumental effects not only distort light curve morphology but also affect spectral measurements, underscoring the importance of their characterization for accurate data analysis.} 
  \label{fig:1}    
\end{figure}

Designed for precise high-energy charged particle measurements, HEPP-H utilizes a multi-component system (DSSD, calorimeter, anti-coincidence detector) to achieve energy/angular detection and particle identification, with sub-microsecond timing \citep{2019RDTM....3...22L, 2019EGUGA..21.3809C}. However, like many high-performance instruments, HEPP-H faces signal accumulation, dead time and other detector response issues during extreme events. Our analysis of high-energy transient sources reveals that exceptionally high count rates induce complex distortions in the observational data, necessitating careful correction for accurate spectral analysis. These effects cause significant light curve distortions (exceeding 50\% amplitude reduction during peak intervals, as shown in Figure~\ref{fig:1}), substantially limiting the scientific utility of HEPP-H data. To address the aforementioned issues, which are inherent to high-rate signal processing, we propose a dynamic hierarchical data acquisition correction method based on MC simulations. This principled approach, which represents a practical implementation of what is known in the field as a ``detector-effects engine'' or ``instrument-effects engine'', constructs a spectral correction function, denoted as $C(E)$, to compensate for the specific data distortions encountered. By applying this correction, we aim to more accurately reconstruct the true incident particle spectrum. This methodology provides a general framework for mitigating systematic uncertainties in high count rate spectroscopy, applicable to particle detectors operating under high count rate conditions where instrumental processing capabilities are significantly impacted, but full saturation has not yet been reached.  

\subsection{Theoretical Framework} \label{subsec:theoretical-foundation}

MC simulations provide a robust theoretical framework for quantifying the intricate interplay of physical processes inherent in detector data acquisition \citep{lepy2019benchmark, 2024arXiv241102606B, 2024arXiv241013402L}. The validity of this approach rests upon three fundamental assumptions: First, the physical processes within the detector system, encompassing incident particle sampling, energy deposition and conversion, signal shaping, and data acquisition chain logic, are assumed to be both decomposable and deterministically parameterizable \citep{brooks1998markov, 2024A&A...691A.319E}. For instance, the Poissonian nature of inter-arrival times can be described by an exponential distribution \citep{grandell1997mixed, grimmett2020probability}, and the temporal convolution of pile-up can be effectively modeled using a Landau function \citep{landau1961prolate, roessl2016fourier}. Second, the inherent measurement bias is essentially a non-linear mapping performed by the detector system on the true physical quantities. This mapping is statistically reversible, enabling the decoupling of pile-up effects, data acquisition system characteristics, and detector response through the construction of a transfer matrix relating true-to-observed energy spectra \citep{craig1986inverse, he2016inversion}. Finally, data acquisition nonlinearities, such as dead time, can be accurately modeled and reproduced. The selection effects introduced by these nonlinearities can then be statistically corrected through large-sample simulations \citep{hsu1947complete, 2010JGRA..115.0E21G}.

A key theoretical advancement of this method lies in establishing a dynamic hierarchical correction approach based on these MC simulations. At the microscopic scale, the method employs MC techniques to simultaneously sample the joint time-energy distribution, rigorously adhering to the stochastic mappings dictated by detector physics processes, as exemplified by Equation \eqref{eq:PH(E)} \citep{Leo1994}. Conversely, at the macroscopic scale, a correction function, $C(E)$, is statistically constructed, as shown in Equation \eqref{eq:C(E)}, to quantify and compensate for the impact of instrumental effects on the acquired data. 

The theoretical rigor of this method is rooted in a comprehensive simulation pipeline, commencing with the sampling of incident spectra corrected for detector efficiency (i.e., the effective area of HEPP-H; \citet{2019RDTM....3...22L}). This progresses through the stochastic transformation of energy deposition into detector channels and culminates in the simulation of pile-up effects and the data acquisition system \citep{Kim01112000}. Ultimately, the statistical convergence of the correction function is guaranteed by the law of large numbers \citep{hsu1947complete}. This ``divide-and-conquer'' strategy ensures that the correction function inherently encompasses both the static characteristics of the detector's intrinsic response and the nonlinear distortion arising under dynamic count rates.

\subsection{HEPP-H Simulation Model} \label{subsec:Simulation}
\subsubsection{MC Simulation of Events and Pile-up}\label{subsubsec:MC pileup}
\begin{figure*}[ht!]  
    \centering
    \begin{subfigure}{0.48\textwidth}
        \includegraphics[width=\linewidth]{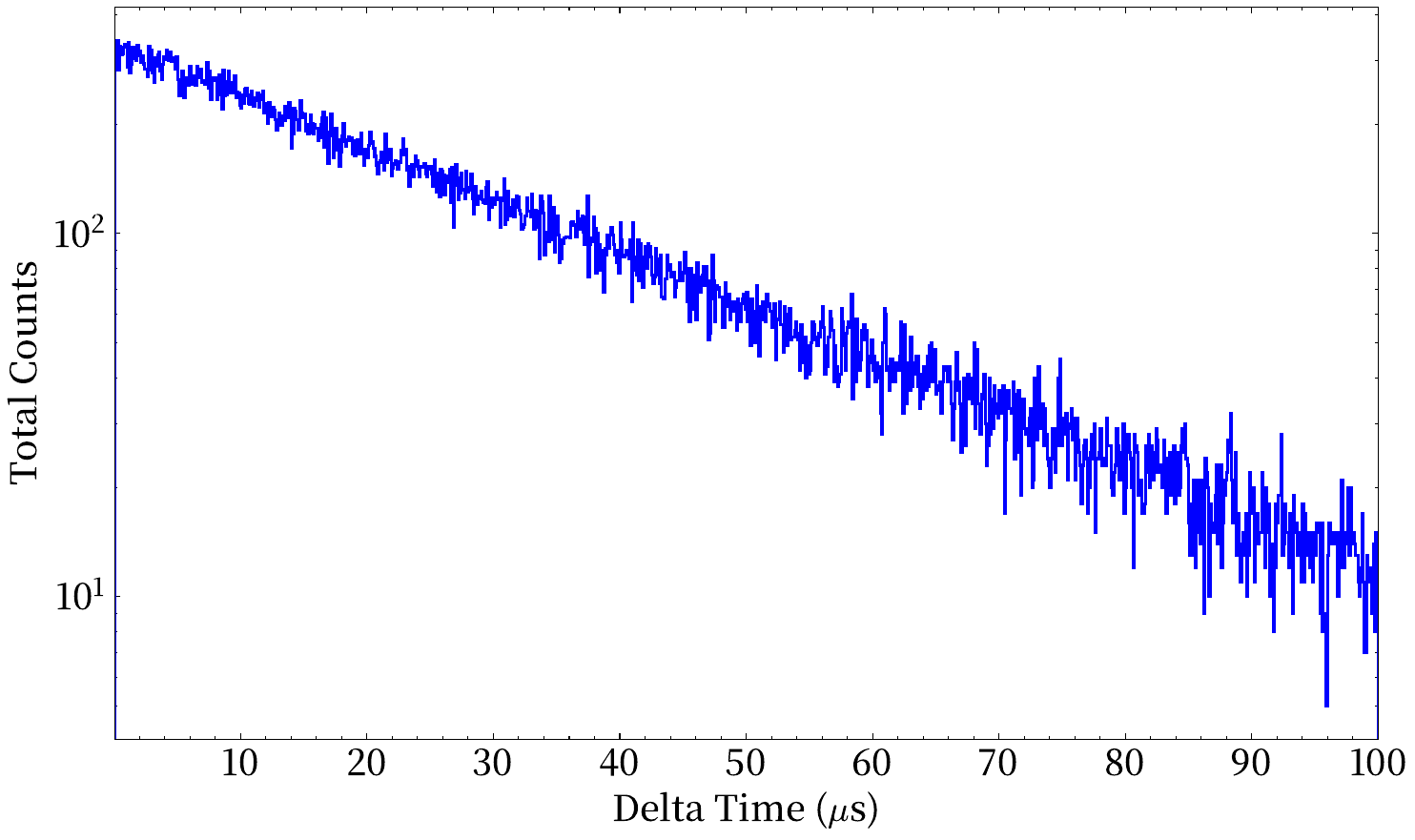}
        \label{fig:time}
    \end{subfigure}\hfill
    \begin{subfigure}{0.48\textwidth}
        \includegraphics[width=\linewidth]{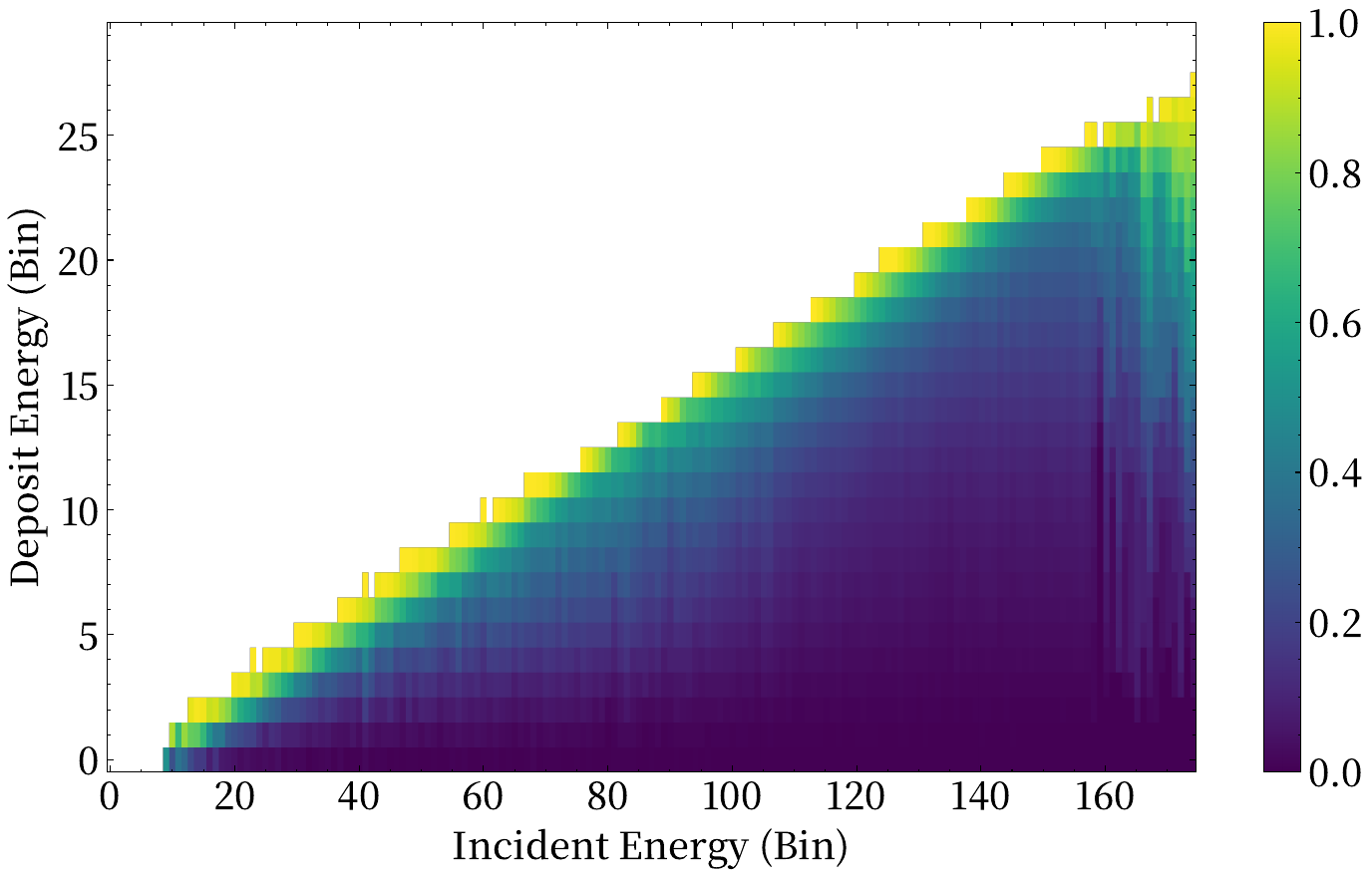}
        \label{fig:RMF}
    \end{subfigure}
    \caption{ \textbf{\textit{Left Panel:}} Distribution of photon arrival time intervals, modeled with an exponential distribution (mean interval of 30~$\mu$s, equivalent to a count rate of 33000~s$^{-1}$). 
    \textbf{\textit{Right Panel:}} Cumulative distribution function of the detector energy response (1.7–600~MeV), calculated using the RMF. This RMF was generated through simulations sampling a uniform incident photon energy spectrum and characterizes the expected detector performance.}
    \label{fig:2}
\end{figure*}  

\begin{figure*}[ht!]  
    \centering
    \begin{subfigure}{0.48\textwidth}
        \includegraphics[width=\linewidth]{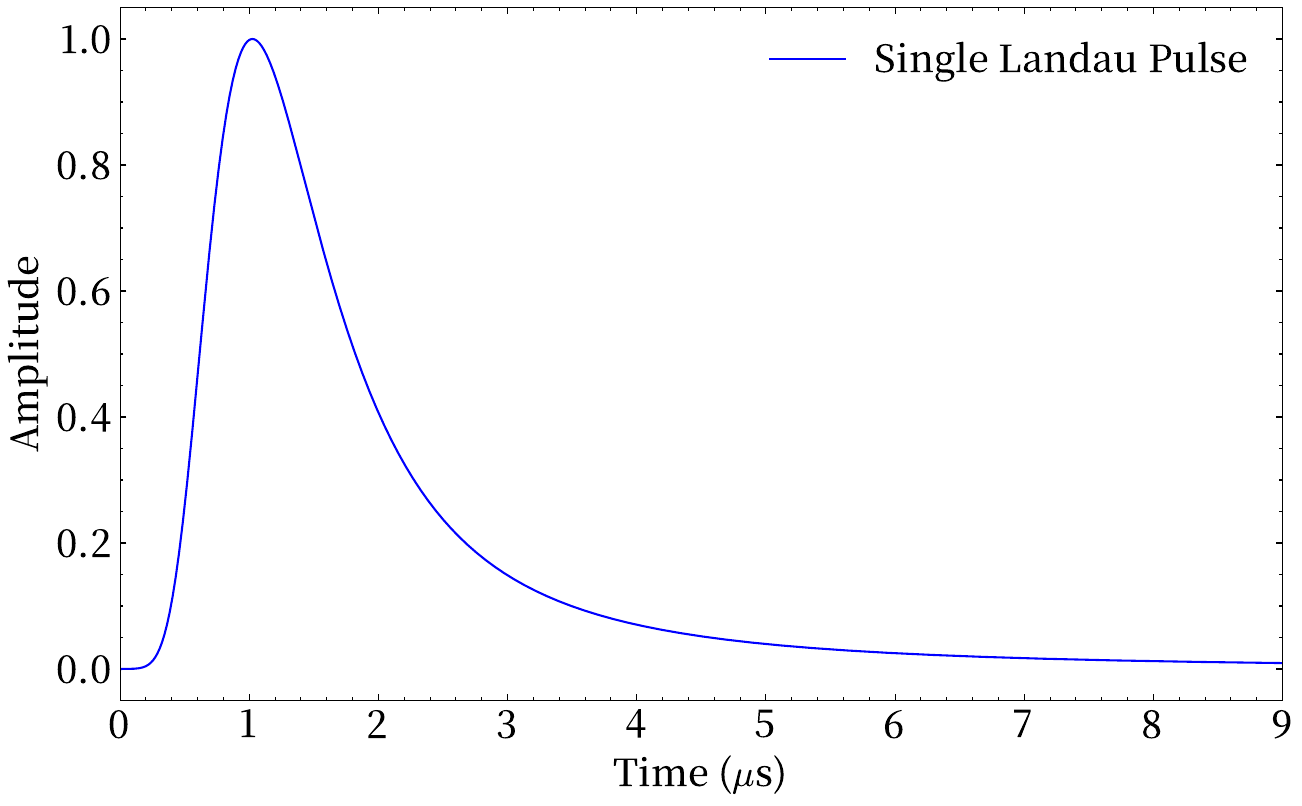}
        \label{fig:landau}
    \end{subfigure}\hfill
    \begin{subfigure}{0.48\textwidth}
        \includegraphics[width=\linewidth]{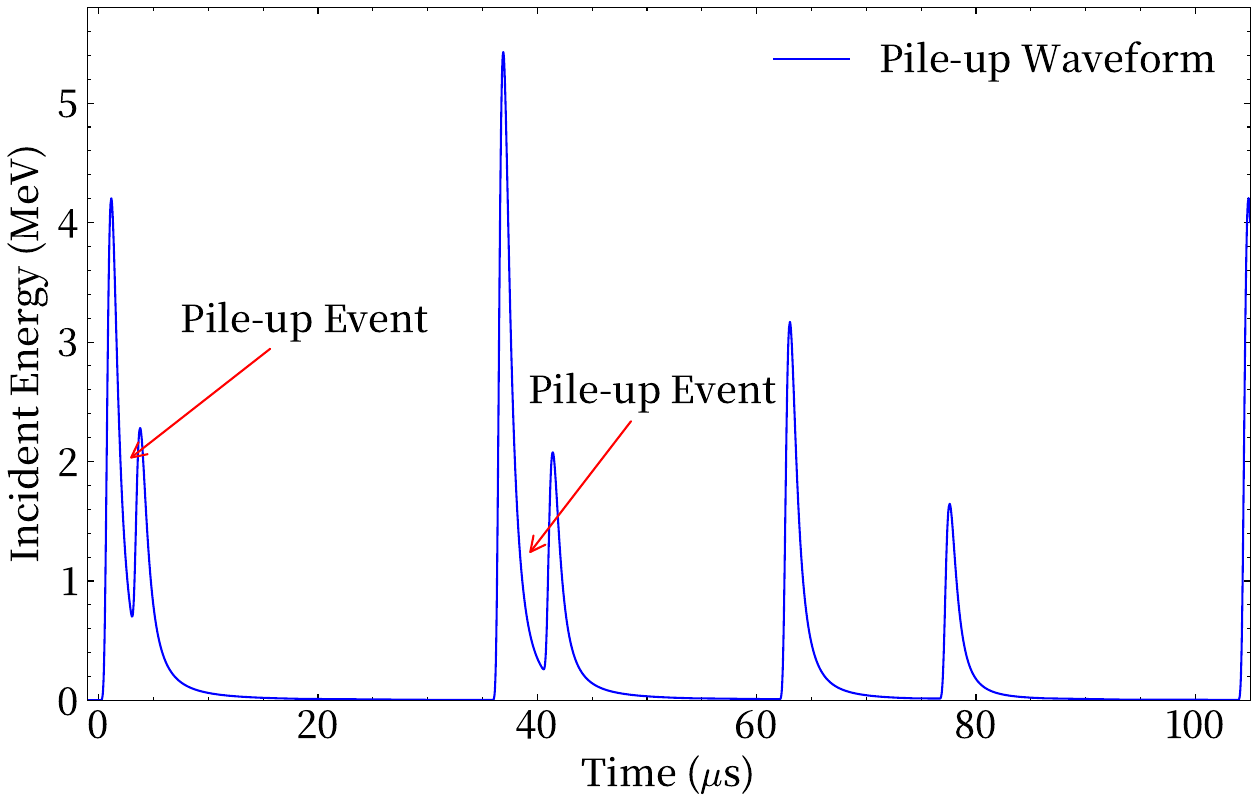}
        \label{fig:pile-up}
    \end{subfigure}
    \caption{Illustrative examples of pulse waveform modeling and pileup simulation. \textbf{\textit{Left Panel:}} Single-event pulse waveform modeled with a Landau distribution, representing a 1~\text{keV} electron-equivalent energy deposition, exhibiting a characteristic rise time of $<$1~$\mu$s and a decay time of 3~$\mu$s. \textbf{\textit{Right Panel:}} Simulated pulse pile-up generated from the linear superposition of individual waveforms within a [-1, +100]~$\mu$s time window around each event arrival.}
    \label{fig:3}
\end{figure*}  

Initially, it is imperative to develop an MC simulation model that faithfully replicates the response of the HEPP-H detector. A central component of this model is the MC event generation and signal superposition module, which aims to emulate the physical processes associated with GRBs, encompassing detector response and pile-up effects.

The MC event generation process initiates by simulating the incident photon characteristics of the observed GRB. To model the temporal arrival of GRB photons, we employ MC sampling, utilizing a parameterized model to describe the spatio-temporal properties of GRB. This process yields an exponential distribution for photon arrival times \citep{grandell1997mixed, grimmett2020probability}:
\begin{equation}
f(t)=
  \left\{\begin{array}{l}
  \lambda e^{-\lambda t}, \quad t \geq 0 ,
  \\0,  \quad {otherwise},
  \end{array}\right.,
  \label{eq:f(t)}
\end{equation}
where $\lambda$ is the rate parameter (typical range: $1$--$10^4$ photons$\, \cdot\, $cm$^{-2}\, \cdot\, $s$^{-1}$).

Figure~\ref{fig:2}, $left$ illustrates the temporal sequence of a Poisson process, with the mean inter-arrival time set to 30~$\mu\mathrm{s}$, to accurately simulate the stochastic arrival behavior of photons at a high count rate of approximately 33 kHz (33000 counts$\, \cdot\, $s$^{-1}$). This aims to realistically represent the extremely high photon flux during the GRB.

For energy sampling, the MC method was similarly employed, in conjunction with a binary search algorithm \citep{sedgewick2011algorithms}, to sample photon energy from the Band model of GRB \citep{1993ApJ...413..281B}. The spectral model parameters used were derived from typical observational analyses of the GRB~221009A burst event \citep{2023arXiv230301203A, 2024SCPMA..6789511Z}, ensuring that the energy distribution of the sampled incident photons accurately represents the spectral characteristics of genuine GRB. The energy spectrum and the temporal distribution of these incident photons provide the initial inputs for an iterative fitting procedure. During this iterative process, a goodness-of-fit  statistic is progressively minimized, with optimal convergence occurring as the modeled spectrum approaches the underlying true incident spectrum. The Band spectral model is shown in equation \citep{1993ApJ...413..281B}:
\begin{equation}
\scriptsize
N(E)=
\left\{\begin{array}{ll}
A\left(\frac{E}{E_{\text {piv }}}\right)^{\alpha} \exp \left(-\frac{E}{E_{\mathrm{c}}}\right), & (\alpha-\beta) E_{\mathrm{c}} \geq E,
\\A\left[\frac{(\alpha-\beta) E_{\mathrm{c}}}{E_{\text {piv }}}\right]^{\alpha-\beta} \exp (\beta-\alpha)\left(\frac{E}{E_{\text {piv }}}\right)^{\beta}, & (\alpha-\beta) E_{\mathrm{c}} \leq E,
\\E_{\text {piv }}=100 \mathrm{keV}, &\end{array}\right.,
\end{equation}
where $A$ is the normalization amplitude constant (photons$\, \cdot\, $cm$^{-2}\, \cdot\, $s$^{-1}\, \cdot\, $keV$^{-1}$), $\alpha$ is the \textit{low-energy} power-law index and $\beta$ is the \textit{high-energy} power-law index, $E_{\mathrm{c}}$ is the characteristic energy in keV, $E_{\mathrm{piv}}$ is the pivot energy in keV and usually the value is taken as 100~keV.

Efficiency correction is achieved for the simulated spectra through convolution with the HEPP-H effective area function (Ancillary Response File, ARF; \citet{2019RDTM....3...22L, Leo1994}).

To simulate the energy response of the HEPP-H detector, this study uses the efficiency-corrected incident photon energy from MC simulations. This energy is then mapped into energy deposition within the detector crystal by employing a pre-constructed Energy Response Matrix (RMF). This RMF, generated through GEANT4 simulations (analogous to the approach used for GECAM; \citet{2020SSPMA..50l9509G}), discretizes the incident energy range into a series of energy channels. For HEPP-H specifically, the incident energy spectrum is binned into 175 logarithmically spaced channels over this range, while the observed detector response is characterized by 30 channels, also covering the 1.37 MeV to 600 MeV range. To elaborate on the mathematical procedure, the RMF is transformed into a Cumulative Distribution Function (CDF, as depicted in Figure~\ref{fig:2}, $right$). The deposited energy spectrum is then ultimately derived by convolving this CDF with the efficiency-corrected incident energy spectrum. The convolution formula is presented below \citep{Kim01112000}: 
\begin{equation}
    PH(E) = \int S(E') \, R(E, E') \, \mathrm{d}E',\label{eq:PH(E)}
\end{equation}
$R(E, E')$ represents the detector energy response function, quantifying the detector's response to an incident photon with energy $E'$.  $S(E')$ denotes the incident gamma-ray energy spectrum, and $PH(E)$ signifies the resulting deposited energy spectrum within the detector.

The simulation of pile-up effects at high count rates is achieved through the linear superposition of single-event signals in the time domain. This approach necessitates an accurate model for the single-event pulse profile, a critical characteristic of which is ``signal trailing''. This effect refers to the finite decay time of the electronic pulse following an energy deposition. If a subsequent event arrives during this decay period, its pulse superimposes on the tail of the preceding one, leading to a baseline shift and an inaccurate energy measurement. To intrinsically account for this behavior, we model the single-event waveform using a Landau function, as illustrated in Figure~\ref{fig:3}, $left$. The simulated waveform exhibits a rise time of less than 1~$\mu\mathrm{s}$ and a signal decay time of 3~$\mu\mathrm{s}$. Its mathematical form is given by \citep{grimmett2020probability, landau196556}:
\begin{equation}
    f(t; t_0, \gamma) = \frac{1}{\pi \gamma \left[ 1 + \left( \frac{t - t_0}{\gamma} \right)^2 \right]},  
\end{equation}
$t$ denotes time, and $f(t; t_0, \gamma)$ represents the waveform amplitude at time $t$.  The parameter $t_0$ signifies the peak time location of the Landau distribution, corresponding to the event arrival time, while $\gamma$ is the scale parameter determining the pulse waveform width.  In the simulations presented herein, $t$ is derived from the cumulative summation of time interval sampling results, $\gamma$ is set to 300~ns, and $t_0$ is established to be proportional to the energy deposition of the event. This proportionality is implemented to reflect the positive correlation between event energy and signal amplitude (These parameters are representative of the typical response characteristics of the HEPP-H plastic scintillator, CsI(Tl) crystal, and SiPM readout electronics; \citet{2019RDTM....3...22L}).  Simulated single-event signal waveforms are linearly superimposed in the time domain.  The total resulting waveform, $V_{\text{total}}(t)$, can be expressed by the following equation:
\begin{equation}
    V_{\mathrm{total}}(t) = \sum_{k} V_{k}(t - t_{k}),
\end{equation}
$V_{\text{total}}(t)$ represents the total superimposed waveform, and $V_{k}(t - t_{k})$ epresents the waveform of the $k$-th event, whose peak is at $t_k$. To effectively simulate pile-up effects, the waveform superposition range, is limited to 1 $\mu\mathrm{s}$ preceding and 100 $\mu\mathrm{s}$ following the arrival time of each event. This ensures accurate simulation of pile-up effects even under conditions of high event density, such as those corresponding to incident photon count rates up to $\sim$$10^6$ counts$\, \cdot\, $s$^{-1}$. It is noteworthy that the peak incident photon count rate for GRB~221009A, as detected by HEPP-H, was lower than this level. Ultimately, the total superimposed waveform, $V_{\text{total}}(t)$, as illustrated in Figure~\ref{fig:3}, $right$, will serve as the input signal for the subsequent hardware-level simulation of the FPGA data acquisition logic.

\subsubsection{FPGA Simulation}\label{subsubsec:FPGA}

To improve the accuracy of our HEPP-H detector simulations and account for DAQ system nonlinearities, we simulated the event data processing logic and timing formed by signals in FPGA. This approach models the core algorithms of the actual DAQ system through a Finite State Machine (FSM) with four states: IDLE, Peak Search, HOLD, and RESET \citep{maxfield2004design, wirthlin2015high}. The state transitions and corresponding DAQ behaviors are described in detail below.

(1) \textbf{IDLE:} This state monitors the input waveform $V_{\mathrm{total}}(t)$, comparing its amplitude to a 1.7~MeV trigger threshold. While $V_{\mathrm{total}}(t)$ remains below threshold, the system stays in IDLE. When the threshold is exceeded, the FSM transitions to Peak Search to initiate data acquisition.

\begin{figure*}[ht!]  
    \centering
    \begin{subfigure}{0.48\textwidth}
        \includegraphics[width=\linewidth]{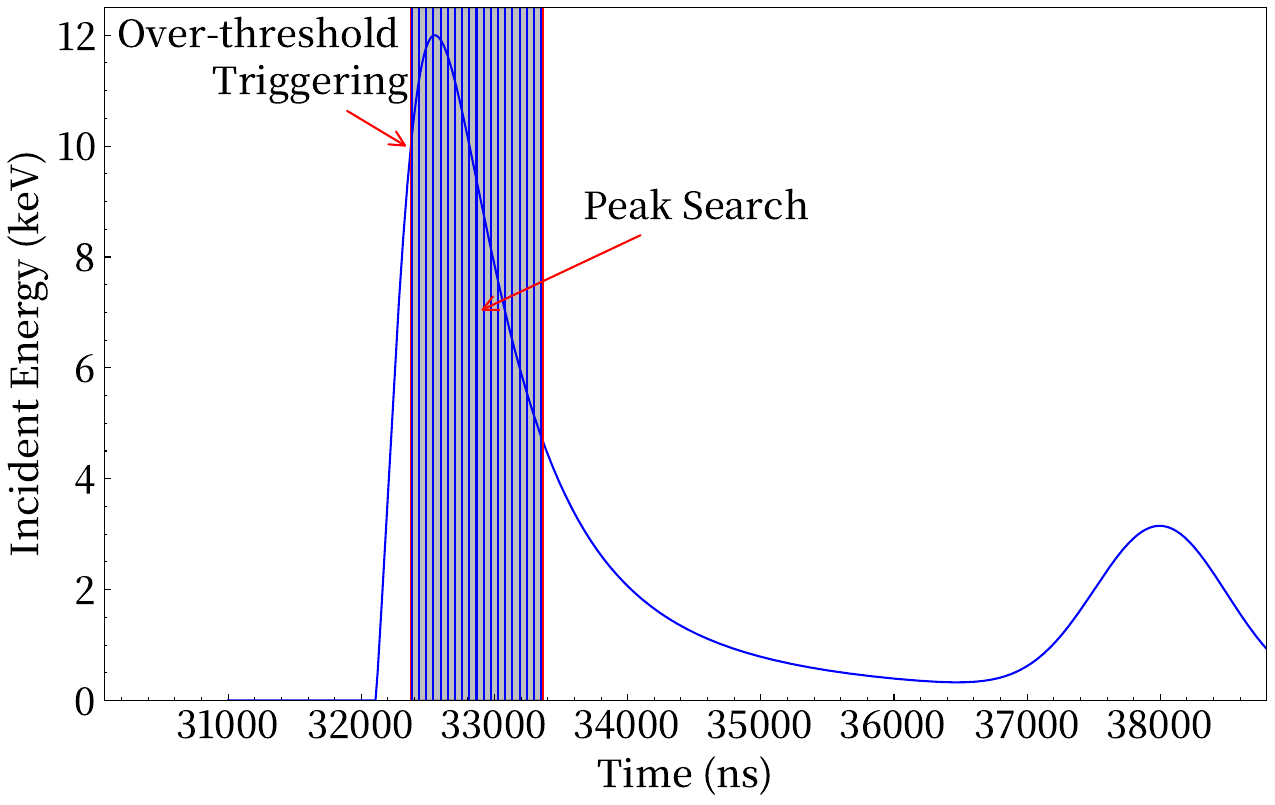}
        \label{fig:Peak search}
    \end{subfigure}\hfill
    \begin{subfigure}{0.48\textwidth}
        \includegraphics[width=\linewidth]{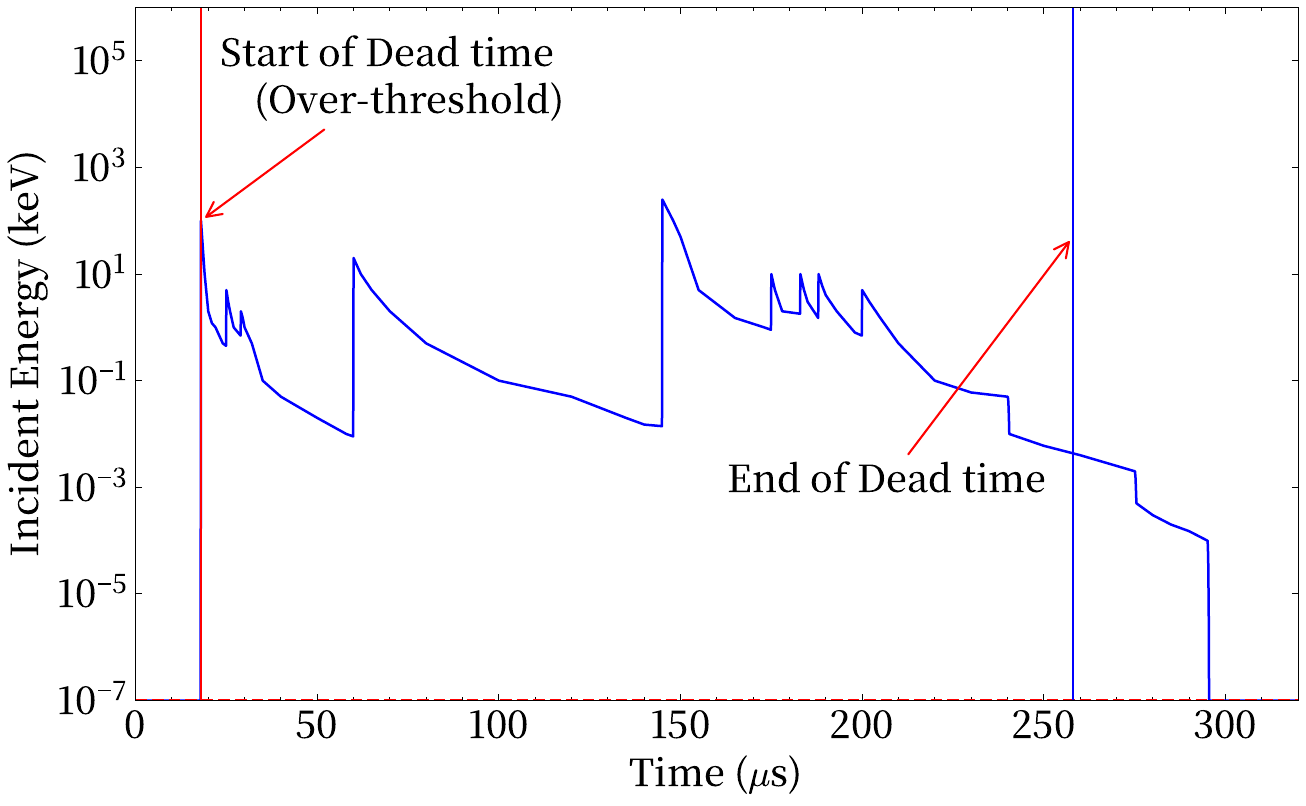}
        \label{fig:Dead time}
    \end{subfigure}
    \caption{ Depiction of the peak search algorithm and dead time implementation. \textbf{\textit{Left Panel:}} Peak detection process, illustrating the 1~$\mu$s detection window (shaded region) initiated when the waveform exceeds the 10~\text{keV} threshold. The ADC sampling rate is 54~ns. \textbf{\textit{Right Panel:}} Illustration of the non-paralyzable dead time. Red and blue vertical lines indicate the onset and termination, respectively, of the dead time period of duration $T_{\mathrm{dead}}$ following an event trigger detection.}
    \label{fig:4}
\end{figure*}

(2) \textbf{Peak Search:}  The DAQ system enters this state to capture the peak amplitude of $V_{\mathrm{total}}(t)$ within a predefined 1 $\mu$s window. During this period, the waveform is sampled at 54 ns intervals through the ADC while maintaining the maximum amplitude value as the \textit{DAQ Energy} measurement (as shown in Figure~\ref{fig:4}, $left$). This 1 $\mu$s window accurately emulates the HEPP-H detector's signal shaping and peak extraction characteristics. The system transitions to \textbf{HOLD} upon peak identification or window expiration to initiate data locking and dead time management. 

(3) \textbf{HOLD:} The system maintains the recorded peak amplitude (DAQ Energy) during a non-paralyzable dead time period of duration $T_{\mathrm{dead}}$. During this period, the DAQ ignores all subsequent triggers regardless of $V_{\mathrm{total}}(t)$ amplitude. As shown in Figure~\ref{fig:4}, right, when the elapsed time since the trigger reaches $T_{\mathrm{dead}}$, the modeled FSM transitions from the \textbf{HOLD} state back to the \textbf{IDLE} state and resumes event monitoring.

(4) \textbf{RESET:} While not explicitly modeled as a distinct state in the current simulation, the core principle of \textbf{RESET} is handled procedurally: following the complete processing of each event, critical event-related variables are cleared, and the system then enters a state of readiness for the subsequent incident photon.

\subsection{Correction Function Construction} \label{subsubsec:C(E)}

\begin{figure*}[ht!]  
    \centering
    \begin{subfigure}{0.48\textwidth}
        \includegraphics[width=\linewidth]{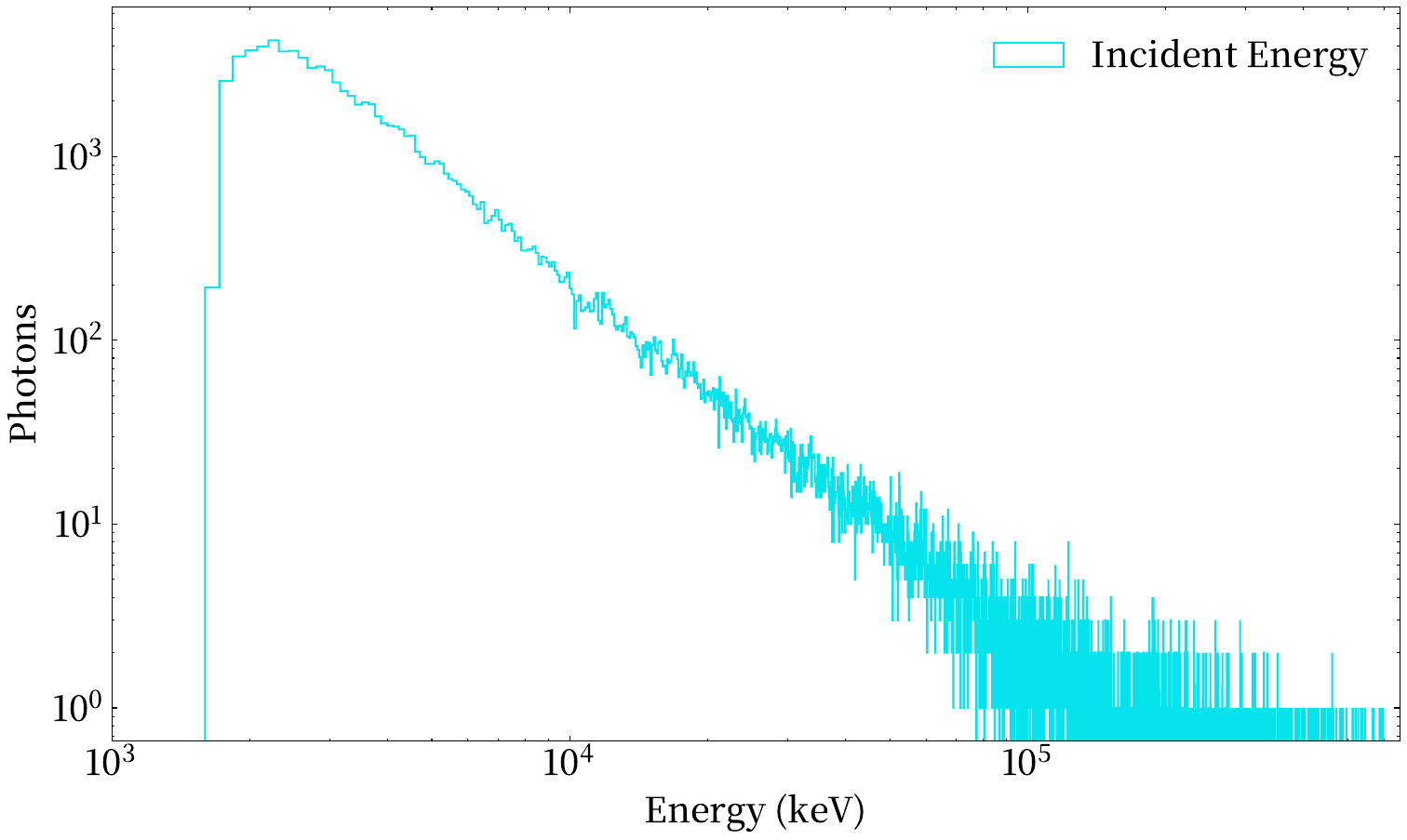} 
        \caption{} 
        \label{fig:top_left}
    \end{subfigure}\hfill 
    \begin{subfigure}{0.48\textwidth}
        \includegraphics[width=\linewidth]{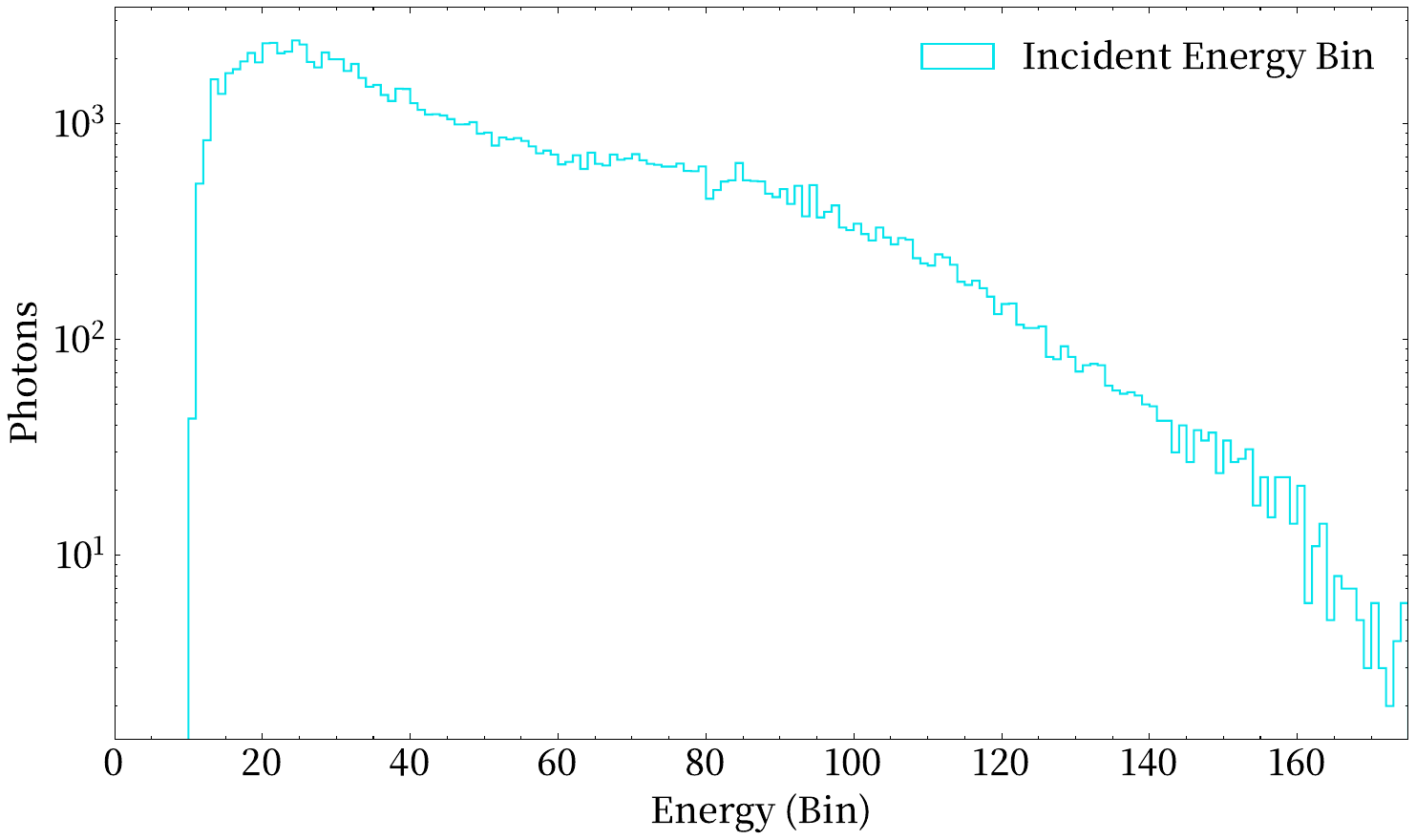} 
        \caption{}
        \label{fig:top_right}
    \end{subfigure}
    \vspace{0mm} 
    \begin{subfigure}{0.48\textwidth}
        \includegraphics[width=\linewidth]{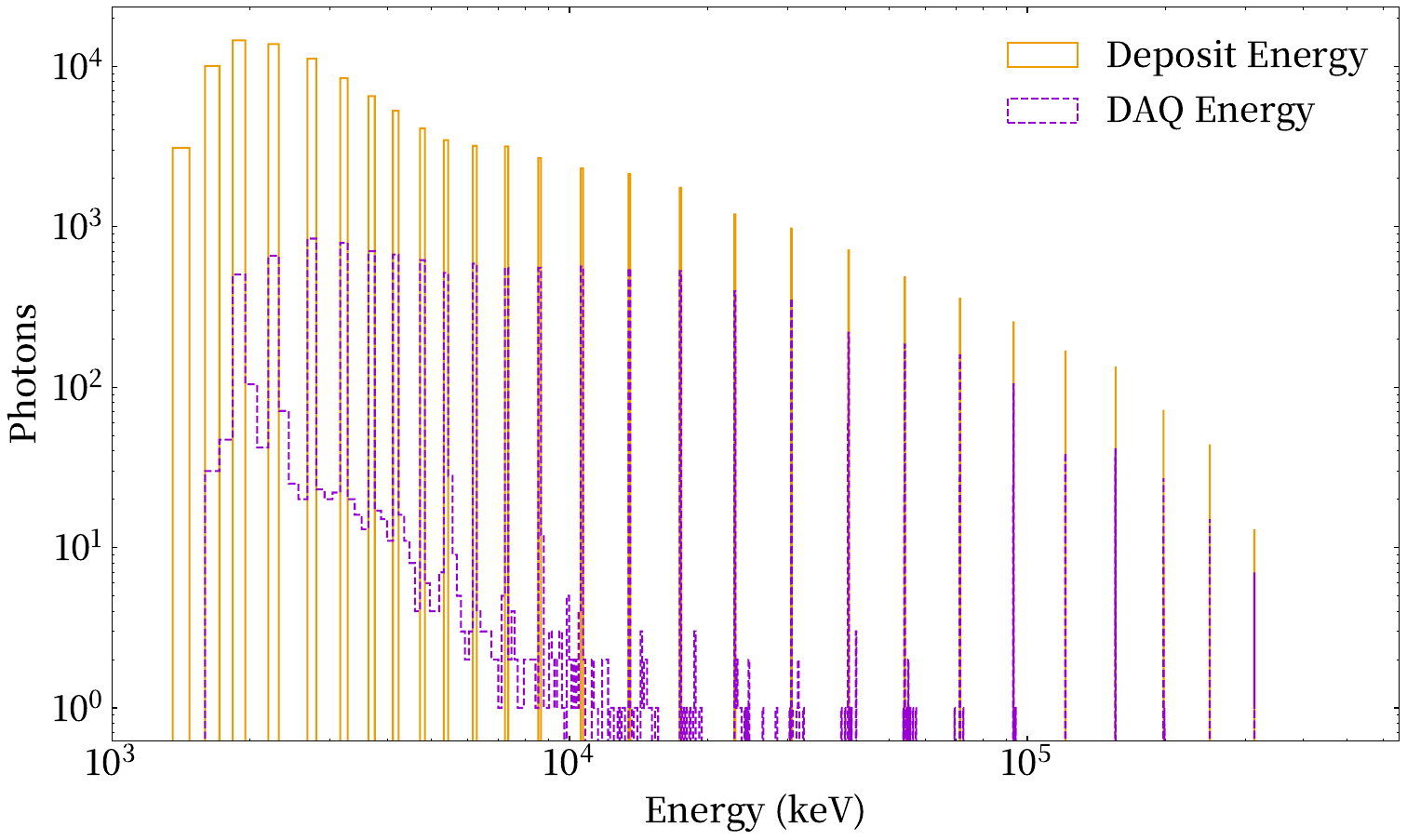} 
        \caption{}
        \label{fig:bottom_left}
    \end{subfigure}\hfill
    \begin{subfigure}{0.48\textwidth}
        \includegraphics[width=\linewidth]{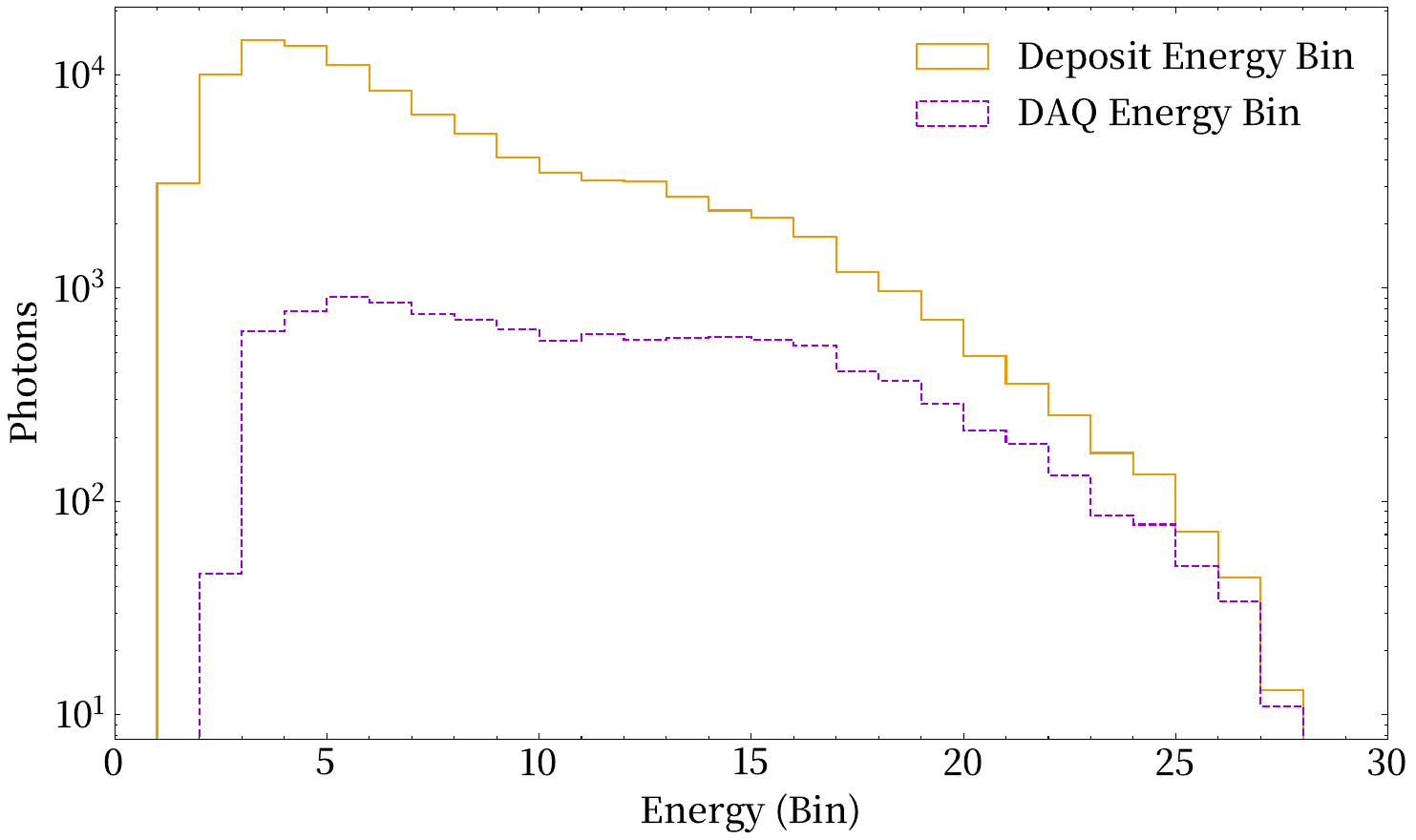} 
        \caption{}
        \label{fig:bottom_right}
    \end{subfigure}

    \caption{Monte Carlo simulation of spectral generation for $10^5$ events. 
    \textbf{(a):} The simulated incident photon spectrum, with photon counts as a function of energy. 
    \textbf{(b):} The same incident spectrum binned into 175 logarithmically spaced energy channels.
    \textbf{(c):} Comparison of the deposited energy spectrum (after applying the Energy Response Matrix to the incident spectrum) and the final Data Acquisition (DAQ) output spectrum (after simulating subsequent instrumental effects). Counts observed between discrete energy channels are attributed to signal pile-up or false triggers. 
    \textbf{(d):} The deposited and DAQ output spectra binned into 30 logarithmically spaced energy channels.
    All simulations used a $1.7$~MeV trigger threshold, a $1\,\mu$s shaping time, and a $240\,\mu$s dead time, with the energy channels covering the 1.37--600~MeV range.}
    \label{fig:5} 
\end{figure*}

The core of our methodology is the construction of an energy-dependent correction function, $C(E)$, derived directly from our Monte Carlo simulations as illustrated in Figure~\ref{fig:5}. The process begins with a simulated incident photon spectrum (panel a), which is then binned to the instrument's channel resolution (panel b). This incident spectrum serves as the input for two parallel simulation paths: one yielding the ideal deposited energy spectrum, $N_{\mathrm{Dep}}$, and the other yielding the final DAQ output spectrum, $N_{\mathrm{DAQ}}$, which includes all instrumental effects. A direct comparison of these two outcomes (panels c and d) reveals the significant spectral distortion introduced by the DAQ process. The correction function is then defined as the energy-dependent ratio of these two spectra. For each energy bin $i$, the discrete correction factor $C(E_i)$ is given by:
\begin{equation}
C(E_i) = \frac{N_{\mathrm{DAQ}}(E_i)}{N_{\mathrm{Dep}}(E_i)} \quad (0 < C(E_i) \leq 1),\label{eq:C(E)}
\end{equation}
where $N_{\mathrm{DAQ}}(E_i)$ is the spectrum distorted by instrumental effects,  while $N_{\mathrm{Dep}}(E_i)$ represents the ideal energy deposition. This ratio quantifies the spectral distortion caused by instrumental effects. By applying smoothing and interpolation to the discrete $C(E_i)$ values, we construct a continuous correction function $C(E)$. This function enables precise compensation of instrumental effects across the entire energy range.

For the analysis of observational data, we first apply the spectral correction function $C(E)$ to the measured counts spectrum. Following this correction, spectral analysis (e.g., deconvolution or model fitting) utilizes the instrument's energy response characteristics. This response is described by RMF, denoted as $R(E, E')$ in Equation~\eqref{eq:PH(E)}, which defines the probability distribution of measuring an event energy $E$ given an incident particle energy $E'$. The corrected distribution is then calculated as:
\begin{equation}
G'(E_i) = G(E_i) \cdot C(E_i),  \label{eq:G(E_i)}
\end{equation}
where $C(E_i)$ accounts for instrumental effects. $C(E_i)$ effectively mitigates count losses and spectral distortions induced by instrumental effects. The corrected RMF$'$ is reconstructed by normalizing the $G'(E_i)$ distributions across all energy channels, denoted as $R'(E, E')$, yielding an improved instrument response for scientific analysis.


\section{Inverse Energy Response Matrix Deconvolution} \label{sec:3}
\subsection{The Deconvolution Problem and Its Prerequisites} \label{subsubsec:P and P}

\begin{figure*}[ht!] 
  \centering 
  \includegraphics[width=0.98\textwidth]{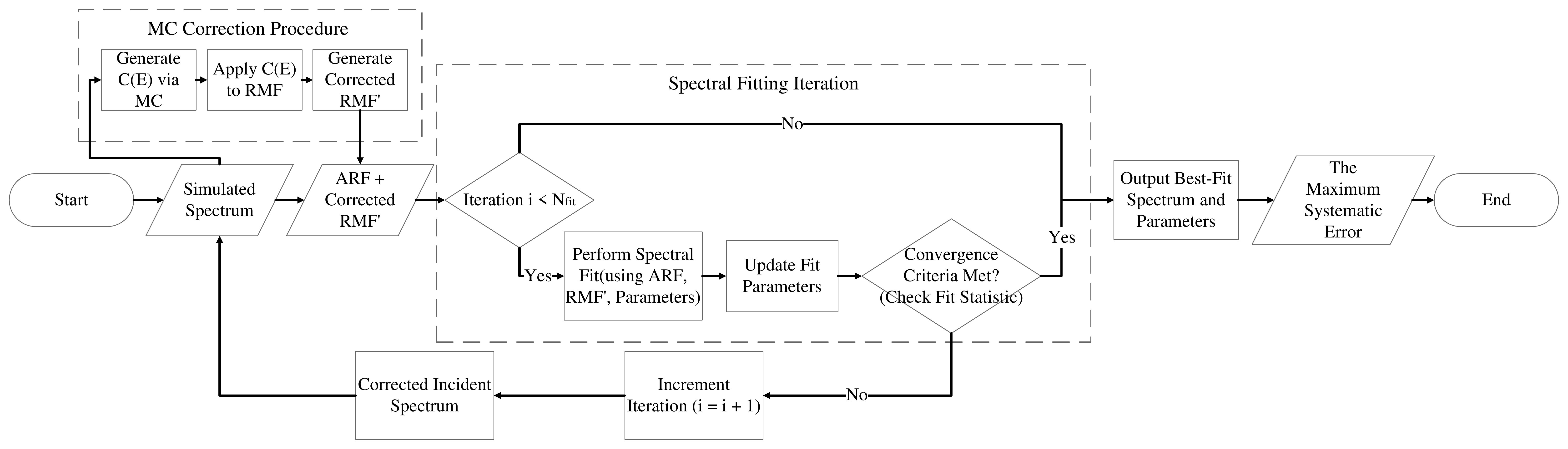} 
  \caption{Flowchart illustrating the iterative spectral fitting procedure incorporating the data acquisition correction function $C(E)$. The process begins with an initial Spectrum used in MC simulations to generate $C(E)$. This function corrects the RMF to produce a corrected response matrix (RMF$'$). An iterative spectral fitting loop (Fit Loop) is then performed using the corrected RMF$'$ and ARF. Within the loop, spectral parameters are fitted to the observed data. If convergence criteria are met or the maximum number of iterations ($N_{fit}$) is reached, the loop terminates, yielding the best-fit spectrum and parameters. A subsequent validation step estimates the maximum systematic uncertainty, particularly that arising from the dependence of the derived $C(E)$ on the initial spectral assumptions and the practical limitations of the iterative process.}
  \label{fig:DFP}    
\end{figure*}

Building upon the data acquisition correction framework (Section~\ref{sec:2}) designed to mitigate instrumental effects, the derivation of the correction function, $C(E)$, involves inherent complexities. The simulation-based calculation of these correction parameters inevitably depends on the assumed incident source spectrum. In the case of the HEPP-H detector, while available rate meter data constrain the total count rate, variations in the assumed input spectral shape and index primarily influence the modeling of low-energy pile-up and the proportion of large-amplitude signals at high energies. Achieving self-consistency thus necessitates an iterative procedure (Figure~\ref{fig:DFP}), cycling between spectral fitting (to refine the incident spectrum) and recalculating the instrumental correction parameters based on this updated spectrum. The core principle of this method is that, ideally, with sufficient iterations in a fully automated simulation-fitting loop, the dependence on the initial assumed spectrum would progressively diminish, converging to a solution characterized by optimal goodness-of-fit metrics (e.g., a reduced $\chi^2_\nu \approx 1$ and a high $p$-value). However, the current practical implementation, which involves manual intervention, inherently reduces iteration efficiency. This necessitates limiting the number of feasible iterations. To address this constraint and to approximate the outcome of a more exhaustive iterative process, we therefore employ a strategy of performing these limited iterations for several distinct initial input spectra and selecting the result that yields the minimum value of the chosen fit statistic.

Therefore, to rigorously quantify the impact of this potential sub-optimality on the derived scientific results, a comprehensive validation process using simulations is essential. We assess the systematic uncertainty introduced by the initial spectral dependence inherent in the $C(E)$ derivation by examining the final output of our complete analysis chain. This involves applying both the $C(E)$ correction and the subsequent inverse energy response matrix deconvolution method (introduced below; Section~\ref{subsubsec:CNN_Inversion}) and comparing the final reconstructed spectrum against the known input spectrum used in the simulation. Such comparison allows for an estimation of the maximum systematic error stemming from the spectral dependence in the $C(E)$ derivation (detailed in Section~\ref{sec:4}).

Once the $C(E)$ correction is applied as part of this analysis, the subsequent, significant challenge is to address the spectral blurring introduced by the detector's intrinsic energy response. Indeed, recovering the incident spectrum accurately---essential for both scientific interpretation and the aforementioned validation of $C(E)$---demands a robust deconvolution technique. Direct spectral inversion is a notoriously ill-posed problem, rendering conventional linear methods inadequate for high-statistics GRB spectra. The failure of such methods is exemplified in Figure~\ref{fig:7}, where a solution derived from Tikhonov regularization is corrupted by severe oscillatory artifacts. This instability arises from a fundamental mismatch between the method's assumptions and the problem's physical reality. Linear regularization techniques like Tikhonov are predicated on the assumption of independent, Gaussian noise. However, high-rate instrumental effects introduce a profoundly non-Gaussian error structure with three key properties. First, the statistical fluctuations themselves are non-Gaussian due to the asymmetric nature of the underlying processes. Second, the response introduces deterministic, non-Gaussian systematic biases, such as spurious pile-up peaks and the non-linear suppression of counts by dead time. Third, and perhaps most critically, the errors in different energy channels become strongly correlated, as pile-up systematically redistributes events from lower to higher energies, thus coupling their statistical fluctuations. The inability of the linear paradigm to account for this multifaceted error structure \citep{hager2015alternating, calvetti2025distributed} necessitates a pivot to a non-linear, data-driven framework.

\begin{figure}[h]
  \centering 
  \includegraphics[width=0.46\textwidth]{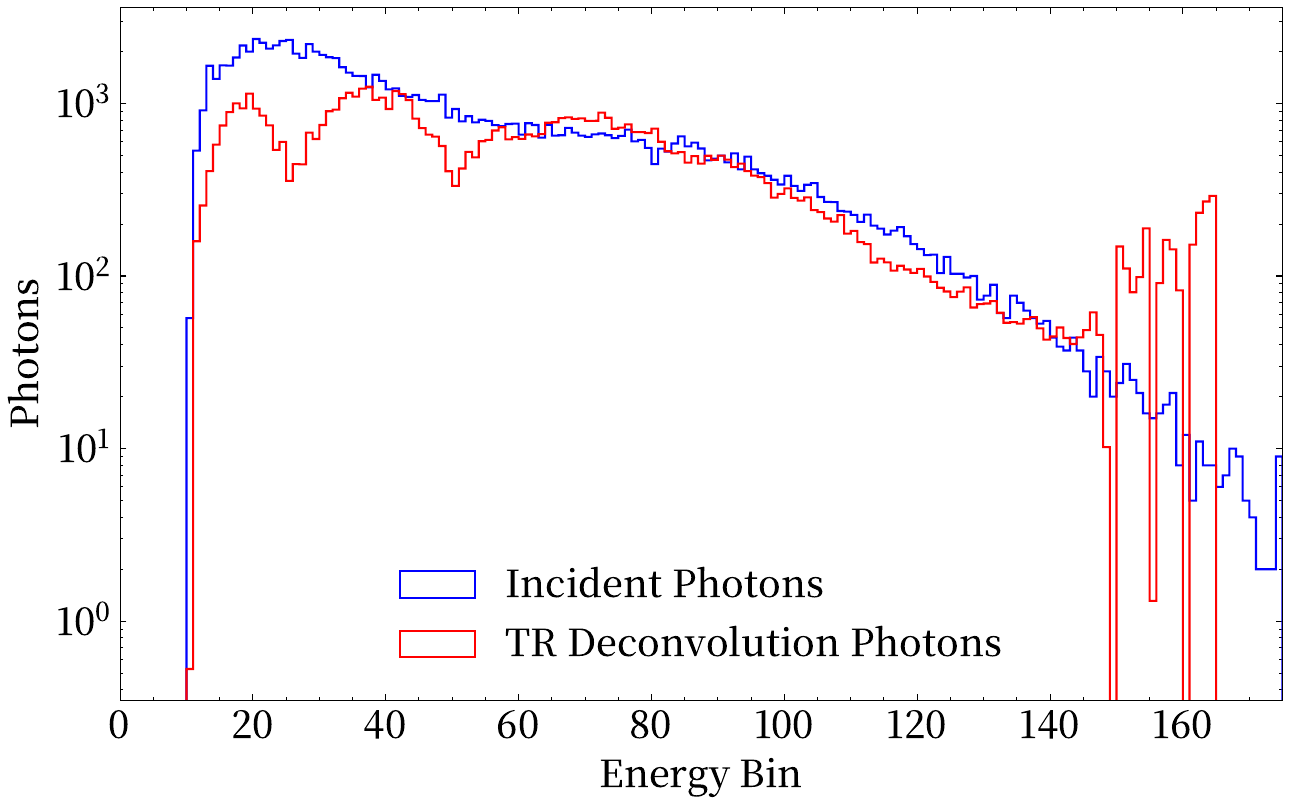}
  \caption{Example of Spectral Inversion via Tikhonov Regularization. This figure illustrates an energy spectrum recovered using Tikhonov regularization, applied to simulated GRB data. The results demonstrate significant sensitivity to noise, evidenced by pronounced oscillations and deviations from the known input spectrum (Input Band spectrum parameters: $\alpha = -1.17$, $\beta = -2.34$, $E_{\mathrm{c}} = 1503~\text{keV}$). The substantial noise amplification highlights a limitation of such regularization methods, particularly when dealing with high count rate spectra.}
  \label{fig:6}    
\end{figure}

To surmount these documented limitations of traditional deconvolution and achieve accurate, model-independent spectral recovery, we introduce an advanced deconvolution method driven by machine learning (ML), specifically employing a Convolutional Neural Network (CNN). Our core innovation lies in training the CNN to implicitly learn and apply the inverse detector response matrix. This approach leverages the CNN's capacity to model complex non-linear mappings effectively and suppress noise, enabling robust and efficient spectral deconvolution. Unlike traditional methods requiring explicit matrix inversion, our CNN-based approach directly predicts the incident spectrum from the corrected detector spectrum through the learned inverse mapping.  
This represents a critical component of the full analysis pipeline, crucial for both scientific analysis and the rigorous validation of our $C(E)$ corrections.

\subsection{CNN-Based Spectral Inversion} \label{subsubsec:CNN_Inversion} 


This learned inverse mapping is implemented by the trained CNN. The mapping is expressed as:
\begin{equation}
\mathcal{F}_\theta: C(E)^{-1}g \rightarrow f_\mathrm{true},
\label{eq:cnn_mapping}
\end{equation}
where $\mathcal{F}_\theta$ denotes the trained CNN acting as the learned inverse operator. The term $C(E)^{-1}g$ represents the data corrected for acquisition effects, and $f_\mathrm{true}$ is the estimated incident spectrum.

Network training incorporates standard practices for stability and robustness \citep{o2015introduction}.
\begin{itemize}
    \item Normalized inputs: $\mathbf{x} = \mathbf{x}/\max(\mathbf{x}_\mathrm{train})$, $\mathbf{y} = \mathbf{y}/\max(\mathbf{y}_\mathrm{train})$ (where $\mathbf{x}$ represents the corrected detector spectrum and $\mathbf{y}$ the corresponding incident spectrum from simulations).
    \item L2 regularization ($\lambda=10^{-3}$) is used along with a Quantile Loss function ($\tau=0.5$):
    \begin{equation}
    L(\mathbf{y}, \mathbf{y}''; \tau) = \frac{1}{N} \sum_{i=1}^N \left[ \tau(y_i - y''_i)_+ + (1-\tau)(y''_i - y_i)_+ \right],
    \end{equation}
\end{itemize}
where $\mathbf{y}''$ is the CNN-predicted incident spectrum. Choosing $\tau=0.5$ (equivalent to Least Absolute Deviation regression) enhances robustness to outliers \citep{arefi2020quantile, 10.1007/978-981-15-0947-6_34}.

\textbf{Inverse Response Matrix Generation and Application}:
The trained CNN, $\mathcal{F}_\theta$, learns a complex mapping from the instrumentally-corrected detector spectrum to the incident spectrum. This learned mapping, developed through training on extensive simulations of diverse incident photon spectra and their corresponding detector responses, inherently captures the inverse characteristics of the detector system. To facilitate direct linear deconvolution and to explicitly represent this learned inverse response, an inverse response matrix, $\mathbf{R}^{-1}_{175 \times 30}$ (mapping 30 detector bins to 175 incident energy bins), is derived from the trained network $\mathcal{F}_\theta$.

This matrix, $\mathbf{R}^{-1}_{175 \times 30}$, effectively embodies the linearized inverse transformation learned by the neural network $\mathcal{F}_\theta$. Built upon the network's ability to generalize from comprehensive training data, it is designed to be a robust and accurate representation for spectral deconvolution. Ideally, this approach provides a method for deconvolving arbitrary observed spectra that is largely independent of specific a priori assumptions regarding the incident spectral model.

Once this inverse response matrix, $\mathbf{R}^{-1}_{175 \times 30}$, is obtained, the incident spectrum is reconstructed by applying it to the $C(E)$-corrected detector spectrum, $\mathbf{x}_{\mathrm{corr}}$, via matrix multiplication:
\begin{equation}
\mathbf{y}'' = \mathbf{R}^{-1}_{175 \times 30} \mathbf{x}'_{\mathrm{corr}},
\label{eq:inversion_apply}
\end{equation}
where $\mathbf{y}''$ is the reconstructed incident spectrum, $\mathbf{R}^{-1}_{175 \times 30}$ is the inverse response matrix, and $\mathbf{x}'_{\mathrm{corr}}$ is the $C(E)$-corrected detector spectrum.

\textbf{Extended-Range Inverse Response Matrix}: To mitigate spectral distortions near the edges of the analysis band caused by energy dispersion, we utilize the CNN $\mathcal{F}_\theta$ to generate an inverse response matrix, $R^{-1}_{\mathbf{N_\text{ext} \times 30}}$, spanning an extended incident energy range ($N_\text{ext} > 175$ channels) broader than the target analysis band ($N_\text{band} = 175$ channels). Generating the matrix over this wider range allows the learned mapping $\mathcal{F}_\theta$ (obtained via Equation~\eqref{eq:cnn_mapping} but with an extended identity matrix conceptual input or trained over the extended range) to better account for photons whose energy is dispersed across the nominal band edges.  

\begin{figure}[h]
  \centering 
  \includegraphics[width=0.46\textwidth]{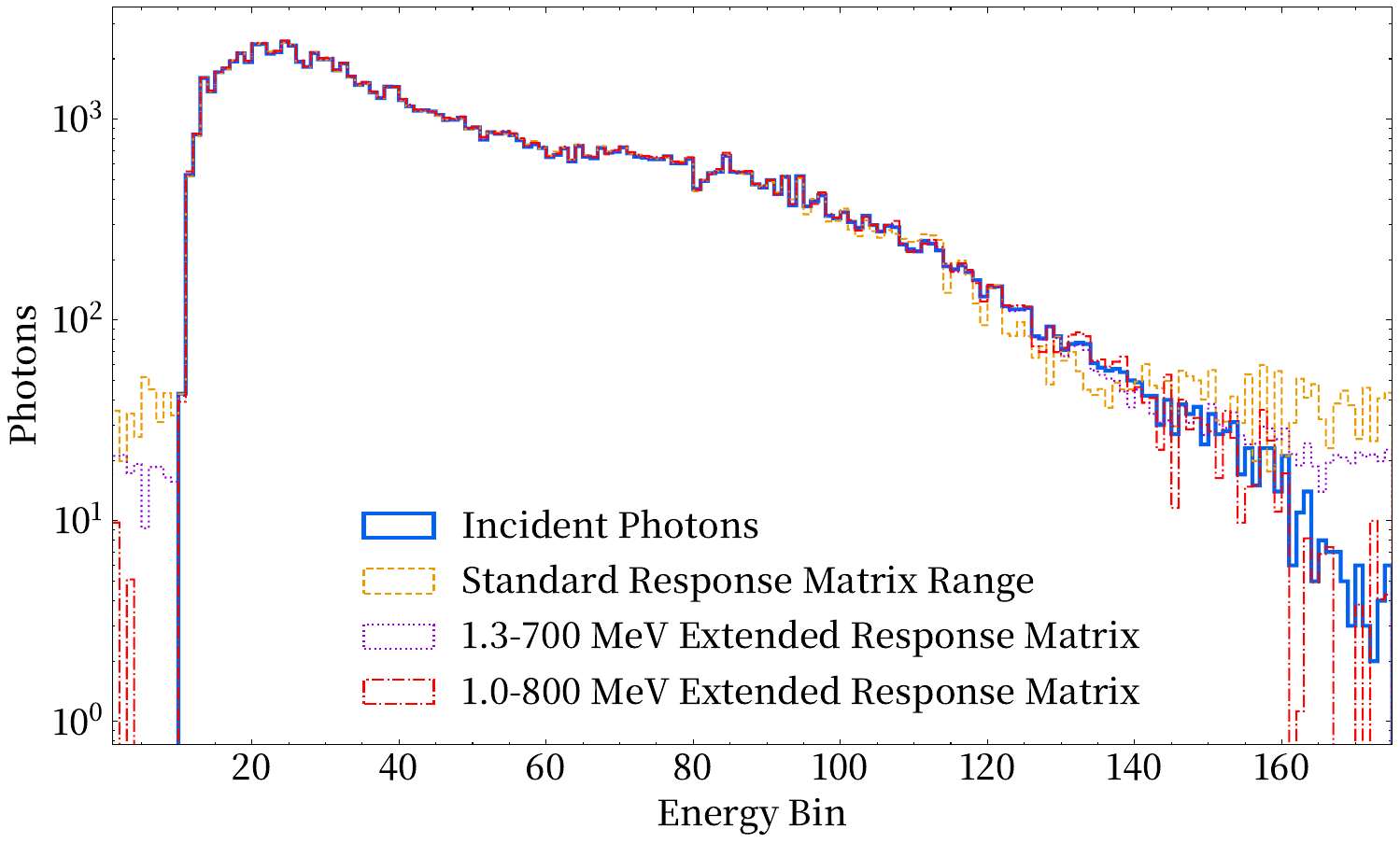}
  \caption{Impact of using an extended energy range during CNN training on spectral inversion accuracy. Compared are the input incident energy spectrum (black) with reconstructed spectra using  CNNs trained with target incident spectra spanning different energy ranges.
  CNNs trained with different energy ranges. The green curve represents reconstruction using a CNN trained only on the nominal 1.7--600~MeV range. The blue and red curves show results from CNNs trained on extended ranges (1.3--700~MeV and 1.0--800~MeV, respectively). Training the CNN to reconstruct spectra over an extended incident energy range demonstrates improved accuracy, particularly minimizing edge distortions.} 
  \label{fig:7}    
\end{figure}

This extended-range matrix $R^{-1}_{\mathbf{N_\text{ext} \times 30}}$ incorporates information about scattering into and out of the primary analysis band more effectively than a matrix generated solely within the $N_\text{band}$ limits. For the final spectral reconstruction within the analysis band, the relevant sub-matrix $R^{-1}_{\mathbf{175 \times 30}}$ is typically extracted from $R^{-1}_{\mathbf{N_\text{ext} \times 30}}$ and then applied to the corrected detector spectrum $\mathbf{x}'_\mathrm{corr}$ using Equation~\eqref{eq:inversion_apply}. This approach minimizes edge distortions and preserves flux accuracy within the analysis band, particularly for spectral features near the band edges (see Figure~\ref{fig:7}).

\begin{figure*}[ht!]  
    \centering
    \begin{subfigure}{0.41\textwidth}
        \includegraphics[width=\linewidth]{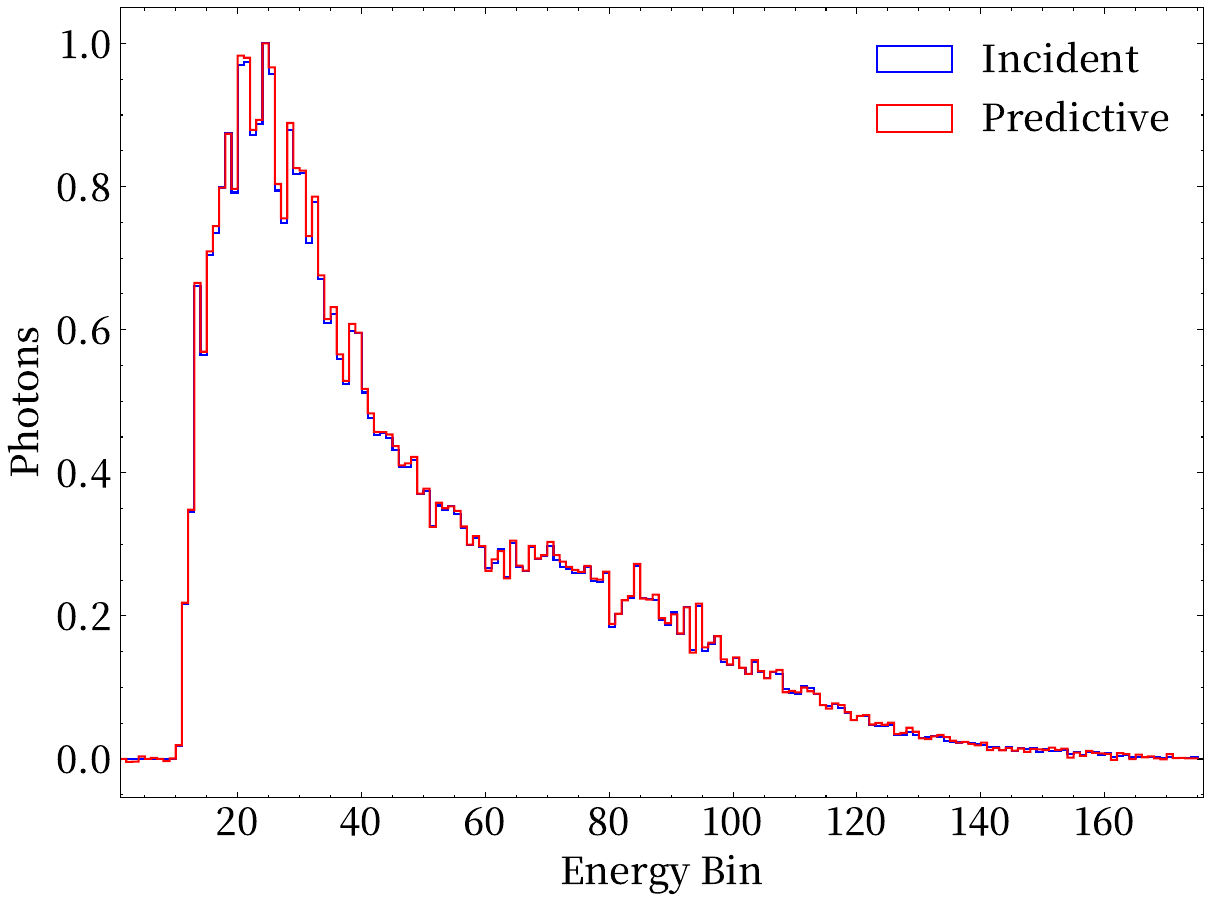}
        \label{fig:Peak search}
    \end{subfigure}\hfill
    \begin{subfigure}{0.48\textwidth}
        \includegraphics[width=\linewidth]{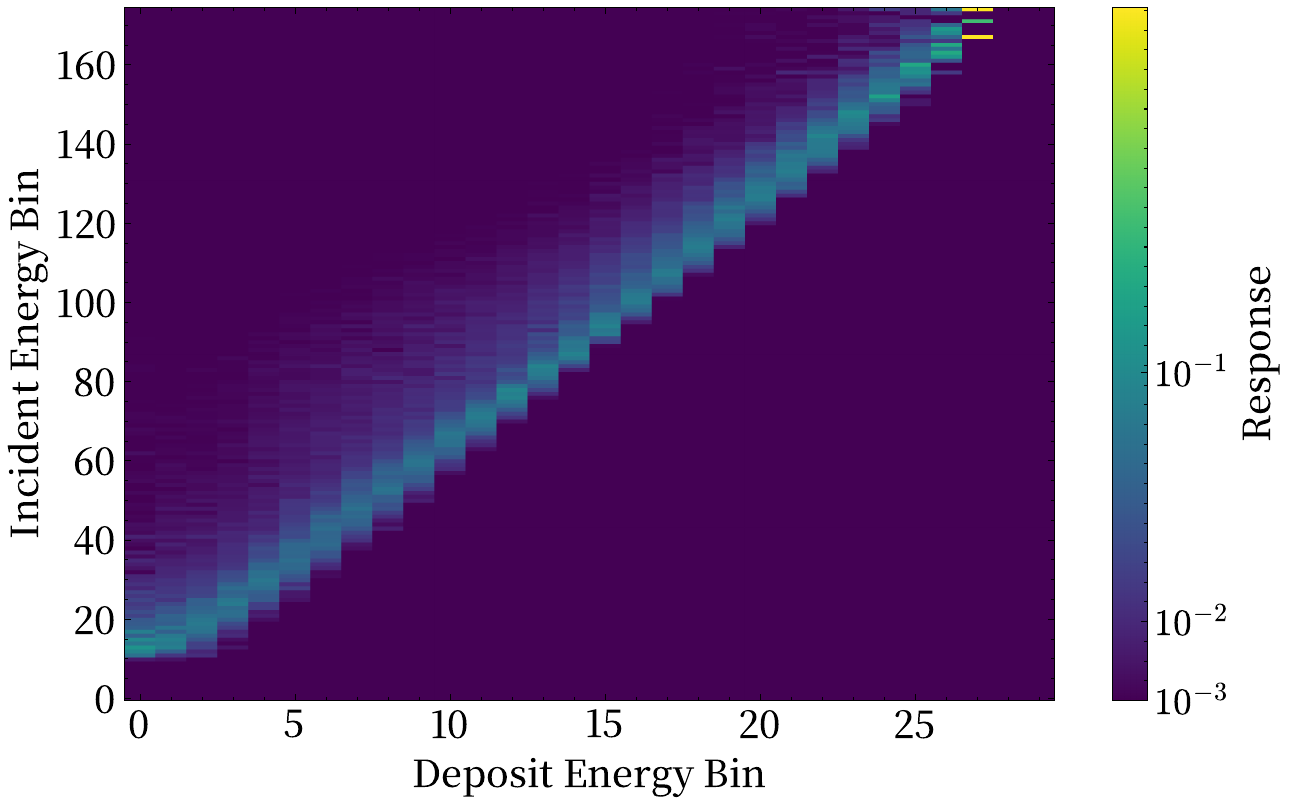}
        \label{fig:Dead time}
    \end{subfigure}
    \caption{\textbf{\textit{Left Panel:}} CNN-predicted energy spectra, comparing the original incident spectrum (blue) with the predicted spectrum (red). \textbf{\textit{Right Panel:}} Visualization of the statistically-derived inverse response matrix used for linear deconvolution. The plot maps deposited energy channels (x-axis) to incident energy channels (y-axis). This matrix was constructed by statistically correlating a comprehensive set of simulated incident spectra with their corresponding ideal deposited spectra, aiming to produce a robust deconvolution method.} 
    \label{fig:8}
\end{figure*}

Training uses the Adam \citep{kingma2014adam} optimizer for 1000 epochs on Monte Carlo-derived pairs $(\mathbf{x},\mathbf{y})$ to develop the CNN, $\mathcal{F}_\theta$, which learns a complex mapping from the corrected detector spectrum to the incident spectrum. Alongside the direct application of $\mathcal{F}_\theta$, an explicit inverse response matrix is also constructed for linear deconvolution. This matrix is statistically derived by correlating a comprehensive set of simulated incident spectra with their corresponding ideal detector deposition spectra. Conceptually, this statistically-derived matrix approximates the linear component of the inverse transformation that $\mathcal{F}_\theta$ itself learns, as both approaches are fundamentally rooted in extracting statistical relationships from extensive simulation data. Example results comparing the CNN-predicted incident spectra with the original simulated incident spectra, along with a visualization of this statistically-derived inverse response matrix, are shown in Figure~\ref{fig:8}.

This CNN-based approach for spectral inversion offers a novel means to estimate the incident spectrum while minimizing reliance on specific spectral model assumptions during the inversion step itself. By directly learning the mapping from the corrected detector spectrum to the incident spectrum, the method leverages the power of neural networks to handle complex detector responses and suppress noise. This provides enhancements in numerical stability and mapping capabilities compared to traditional inversion techniques, facilitating improved precision in analyzing GRB energy spectra from instruments like HEPP-H.

Our CNN-based spectral inversion method is an example of the broader application of machine learning techniques that continue to transform GRB analysis. Other pioneering studies further illustrate this trend: for instance, \citet{2019Ap&SS.364..105H} developed principal component analysis and clustering techniques for GRB classification using \textit{Fermi}/GBM data, while \citet{2021A&C....3400441M} advanced categorization through fuzzy cluster analysis. Most relevant to our work, \citet{2025ApJS..276...62C} successfully applied CNNs to GRB duration classification, overcoming limitations of traditional methods. These studies collectively highlight the expanding applications of machine learning in GRB research. 

\section{Systematic Error Quantification and Validation} \label{sec:4}

This study rigorously validates the data correction and deconvolution framework through comprehensive Monte Carlo simulations. The validation protocol is designed to quantify the method's performance under both ideal and mismatched conditions, thereby assessing its accuracy, generalizability, and the systematic uncertainties inherent in the process. The validation integrates two key components: a self-consistency test and an extensive cross-validation study using 27 distinct spectral models. 

\subsection{Simulation Framework and Validation Protocol} \label{subsec:SF VP}
\subsubsection{Simulation Data and Test Structure} \label{subsec:Data Generation}
                                
The entire validation procedure pivots on a single, fixed correction function, $C(E)$, which is derived from a canonical reference spectrum. For this reference, we adopt a Band function model with parameters representative of a bright GRB (hereafter referred to as spectrum $A$; see the bolded row in Table~\ref{tab:2}). Following the simulation framework described in Section~\ref{subsec:Simulation}, we generate the simulated instrument response to this input, yielding the observed spectrum $A'$ and the ideal deposition spectrum $A_{\mathrm{Dep}}$. The correction function $C(E)$ is then uniquely determined by the ratio of $A'$ to $A_{\mathrm{Dep}}$ (Section~\ref{subsubsec:C(E)}). This function, once derived, remains fixed for all subsequent tests.

The second key component of the pipeline is the deconvolution operator, implemented as a CNN. It is crucial to note that this CNN is a universal inversion tool, pre-trained on a vast and diverse set of thousands of simulated spectra, independent of the specific spectra used in these validation tests. This pre-trained CNN model also remains fixed and unchanged throughout the self-consistency and cross-validation procedures. 

With this single $C(E)$ established, we perform two distinct tests to evaluate the pipeline:

1. Self-Consistency Test: This test assesses performance under ideal conditions. We apply the correction function $C(E)$ and the CNN-based deconvolution back to the simulated data from which it was derived. The objective is to verify that the final reconstructed spectrum, $A''$, accurately recovers the original incident spectrum $A$.

2. Cross-Validation: This series of tests evaluates the method's generalizability under realistic conditions. We apply the same fixed $C(E)$ to 27 different incident spectra (hereafter referred to as spectra $B$), whose parameters are listed in Table~\ref{tab:1}. For each spectrum $B$, the simulation yields an observed spectrum $B'$, which is then corrected and deconvolved to produce a reconstructed spectrum $B''$. The comparison of each $B''$ to its true counterpart $B$ (with results for all 27 tests compiled in Table~\ref{tab:2}) quantifies the systematic error introduced when applying a fixed correction function to mismatched spectral morphologies.

\begin{table}[htbp]
\centering
\caption{The brightest GRB~221009A spectral parameter ranges and parameter combinations for cross-validation}
\label{tab:1}
\begin{tabular}{cccc} 
\toprule
\textbf{Parameter} & \textbf{Type} & \textbf{Value/Range} & \textbf{Unit} \\
\midrule
\multicolumn{4}{l}{\textbf{Parameter Ranges}} \\ 
\midrule
$\alpha$ & Range & $[-1.60, -0.56]$ & -- \\
$\beta$ & Range & $[-3.00, -1.87]$ & -- \\
$E_\mathrm{c}$ & Range & $[324, 2431]$ & keV \\
\midrule
\multicolumn{4}{l}{\textbf{Parameter Combinations}} \\ 
\midrule
$\alpha$ & Selected & $-0.66$ & -- \\
$\beta$ & Selected & $-1.90$ & -- \\
$E_\mathrm{c}$ & Selected & $640$ & keV \\
\midrule
$\alpha$ & Selected & $-1.17$ & -- \\
$\beta$ & Selected & $-2.34$ & -- \\
$E_\mathrm{c}$ & Selected & $1503$ & keV \\
\midrule
$\alpha$ & Selected & $-1.50$ & -- \\
$\beta$ & Selected & $-2.85$ & -- \\
$E_\mathrm{c}$ & Selected & $2250$ & keV \\
\bottomrule
\end{tabular}
\end{table}

\subsubsection{Validation Results and Analysis} \label{subsec:SCT CV}

\begin{figure*}[ht!]  
    \centering
    \begin{subfigure}{0.48\textwidth}
        \includegraphics[width=0.9\linewidth]{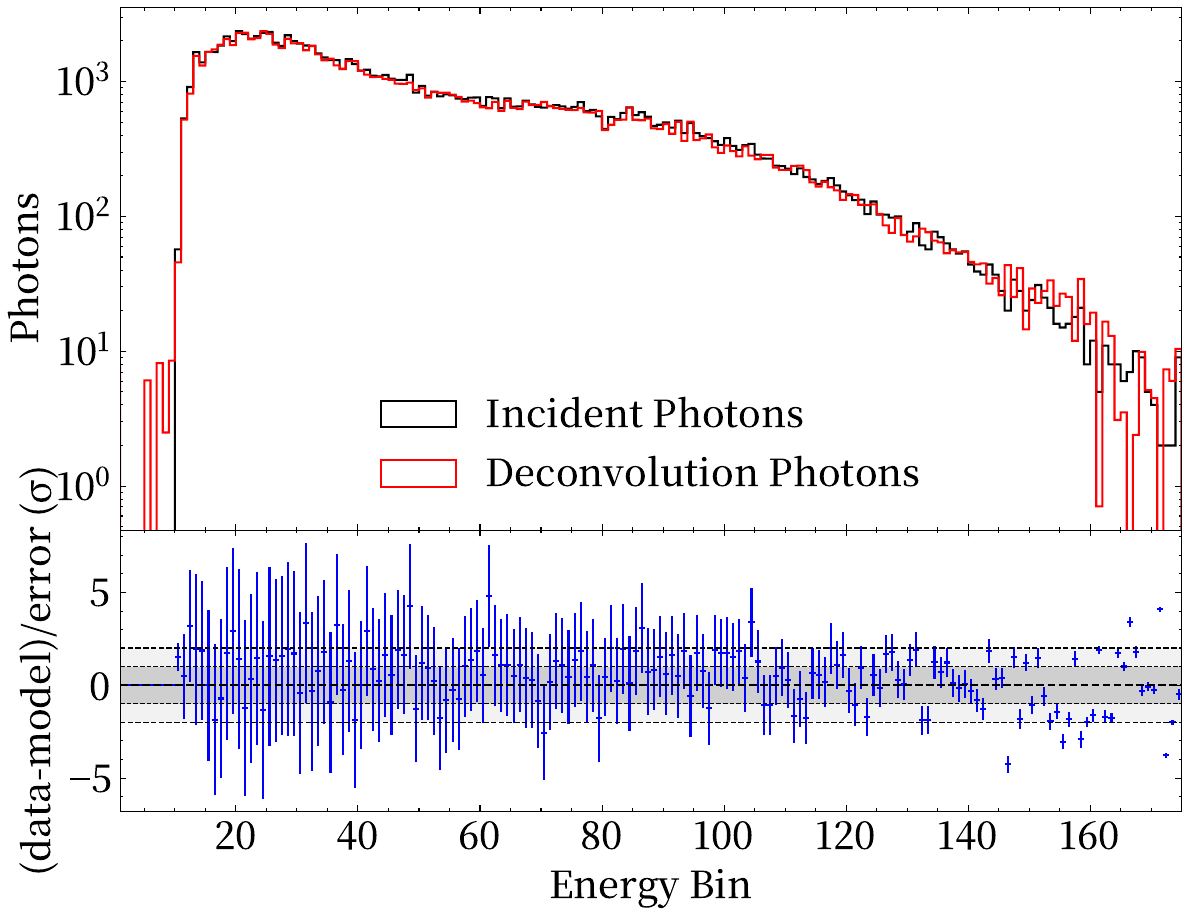}
        \label{fig:Peak search}
    \end{subfigure}\hfill
    \begin{subfigure}{0.48\textwidth}
        \includegraphics[width=0.9\linewidth]{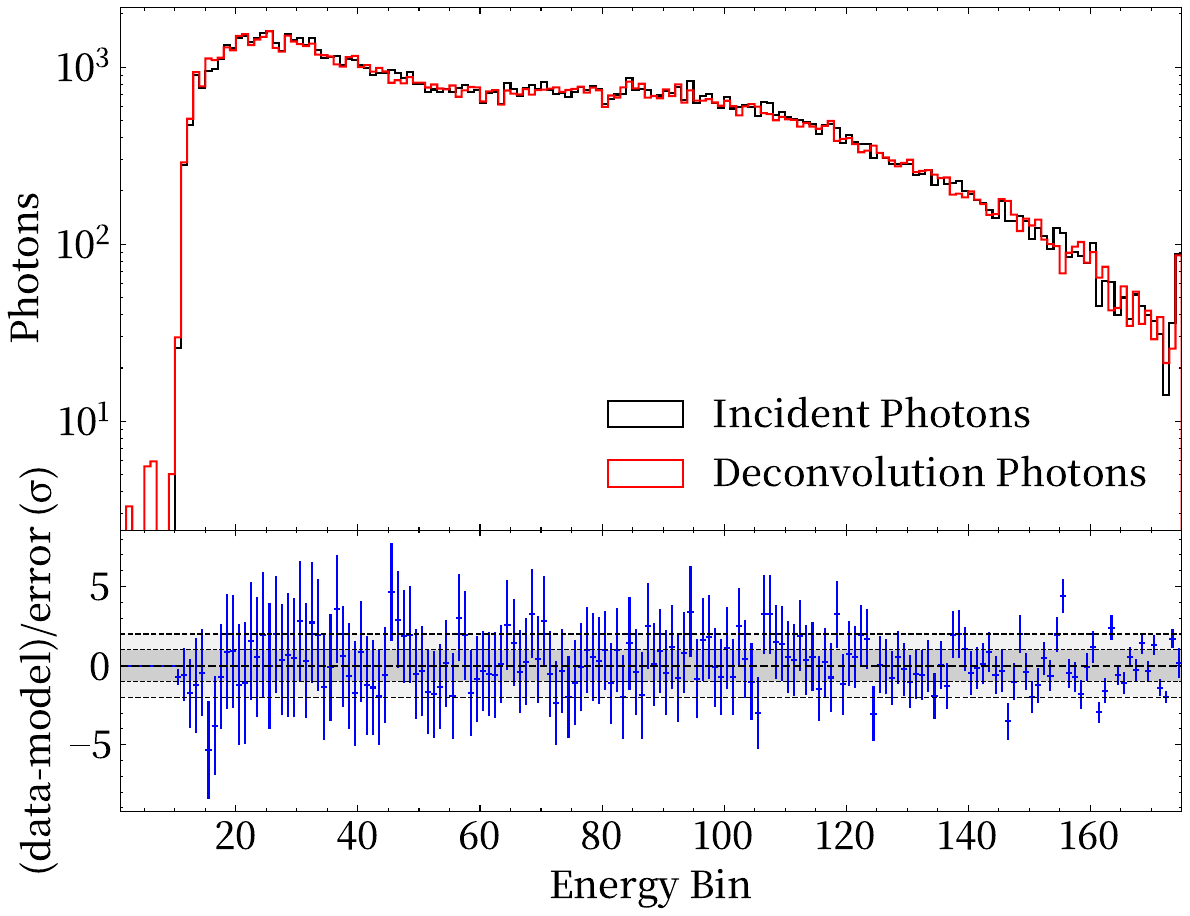}
        \label{fig:Dead time}
    \end{subfigure}
    \caption{Energy spectra after $C(E)$ correction and deconvolution. \textbf{\textit{Left Panel:}} A self-consistency test showing the deconvolved spectrum (input: $\alpha = -1.17$, $\beta = -2.34$, $E_\mathrm{c} = 1503~\text{keV}$). The residuals are centered on zero, with \textbf{90.9\%} of the data points falling within the $\pm$2$\sigma$ bounds. \textbf{\textit{Right Panel:}} A cross-validation test using the same correction function and CNN on a different input spectrum ($\alpha = -0.66$, $\beta = -1.90$, $E_\mathrm{c} = 2250~\text{keV}$). The residuals remain centered on zero, though the distribution is slightly broader, with \textbf{87.3\%} of the points within $\pm$2$\sigma$. This increased spread is expected due to the model mismatch inherent in the cross-validation process.} 
    \label{fig:9}
\end{figure*}

The self-consistency test demonstrates an exceptionally accurate spectral reconstruction, as shown in the left panel of Figure~\ref{fig:9}. A residual analysis confirms the high fidelity of the result. The residuals are symmetrically distributed about zero, a strong indication that the reconstruction is free from significant systematic bias. A quantitative analysis of the residual spread reveals that 90.9\% of the data point centers fall within the $\pm$2$\sigma$ confidence bounds (while 93.1\% of the corresponding error bars overlap with this region). While this fraction deviates from the 95.4\% expected for a purely Gaussian distribution, the wider-than-expected spread is attributable to the additional, non-Gaussian uncertainties introduced by the non-linear deconvolution process. The ideal statistical uncertainty serves as a useful metric, but it does not fully encompass the complex error propagation through the inversion algorithm.

To provide a rigorous, quantitative assessment of these comparisons, we supplemented the descriptive metrics in Table~\ref{tab:2} with two non-parametric goodness-of-fit tests: the Kolmogorov-Smirnov (K-S) test \citep{press2002numerical} and the more tail-sensitive Anderson-Darling (A-D) test \citep{anderson1952asymptotic}. Both tests evaluate the null hypothesis ($H_0$) that the reconstructed spectrum and the incident spectrum are samples drawn from the same underlying parent distribution. Following convention, a p-value greater than the standard significance level of 0.05 indicates that we fail to reject this null hypothesis, implying no statistically significant difference between the two spectra \citep{Fisher1925}.

The results reveal a nuanced picture of the method's performance. In the ideal self-consistency test (bolded row in Table~\ref{tab:2}), both the K-S test ($p=0.429$) and the A-D test ($p>0.25$) yield p-values well above the $0.05$ significance level, confirming the core algorithm's accuracy. In the more stringent cross-validation tests, the K-S test passes in 23 of 27 cases (85\%), indicating that the overall spectral shape is robustly recovered. The A-D test, however, passes in only 15 cases (56\%). The failures are predominantly for spectra whose parameters deviate most from the reference model. This indicates that while the bulk of the spectrum is reconstructed with high fidelity, the model mismatch inherent in cross-validation can introduce minor, statistically significant deviations in the spectral tails, which are detected by the sensitive A-D test.

\subsection{Systematic Error Analysis} \label{subsec:S E}
\begin{figure}[ht!]
    \centering
    \begin{subfigure}{0.45\textwidth} 
        \includegraphics[width=\linewidth]{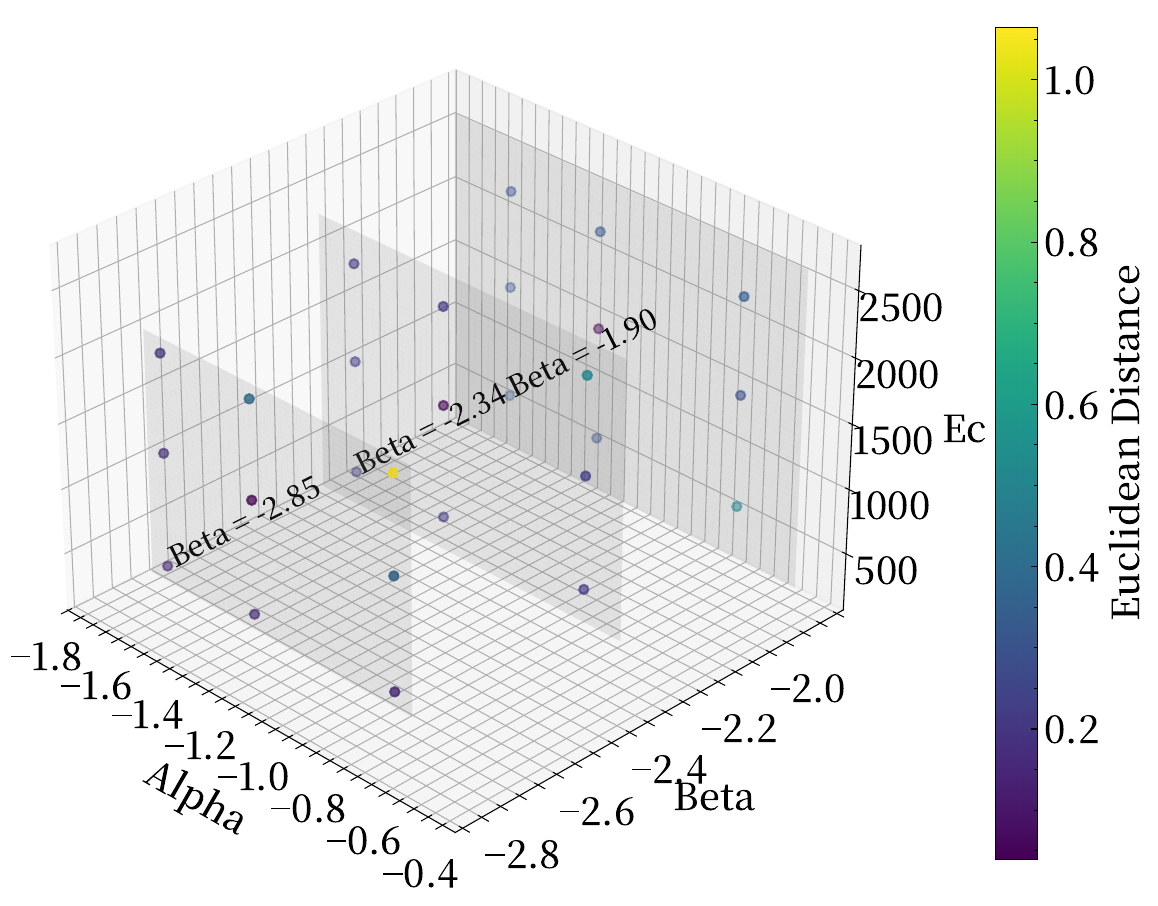}
        \caption{Parameter space correlations} 
        \label{fig:subfig1} 
    \end{subfigure}
    \hfill 
    \begin{subfigure}{0.45\textwidth} 
        \includegraphics[width=\linewidth]{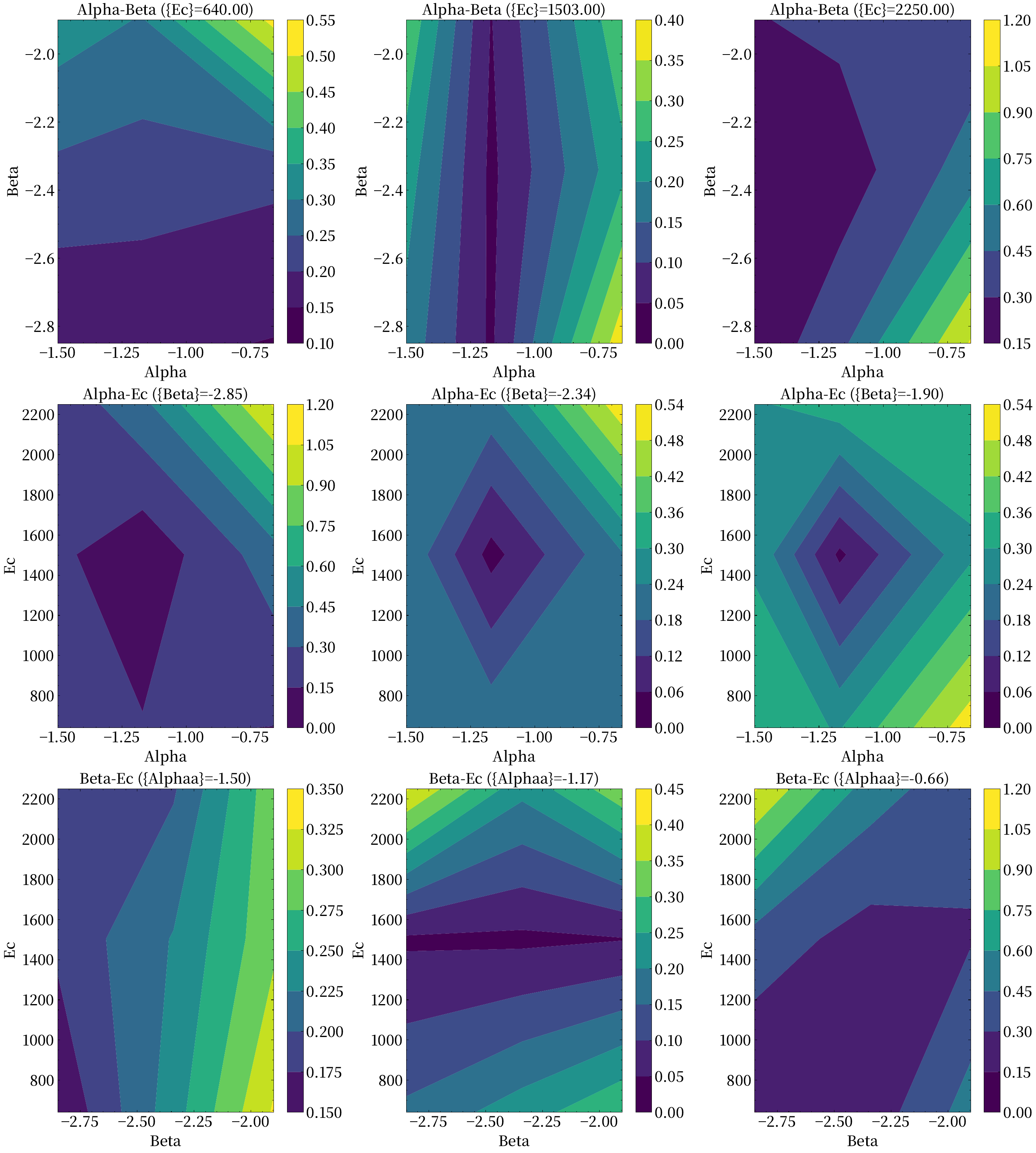} 
        \caption{Contour analysis with different parameters fixed} 
        \label{fig:subfig2} 
    \end{subfigure}
    \caption{\textbf{{(a):}} Three-dimensional parameter space visualization showing Euclidean distance (color-coded) between corrected and incident spectra for 27 spectral configurations. Axes correspond to $\alpha$ (range: $-1.8$ to $-0.40$), $\beta$ (range: $-3.0$ to $-1.6$), and $E_c$ (range: $300$ to $2500$ keV). 
    \textbf{{(b):}} Nine subplots showing parameter parsimony for fixed $E_c$ (top row), fixed $\beta$ (middle row) and fixed $\alpha$ (bottom row) conditions. All contours were calculated based on Euclidean distances from the 27 sets of energy spectra. Characteristic parameters are chosen as $E_c \in \{640,1503,2250\}$ keV, $\beta \in \{-2.85,-2.34,-1.90\}$, $\alpha \in \{-1.50,-1.17,-0.66\}$.
    } 
    \label{fig:10} 
\end{figure}

\begin{figure}[h] 
  \centering 
  \includegraphics[width=0.46\textwidth]{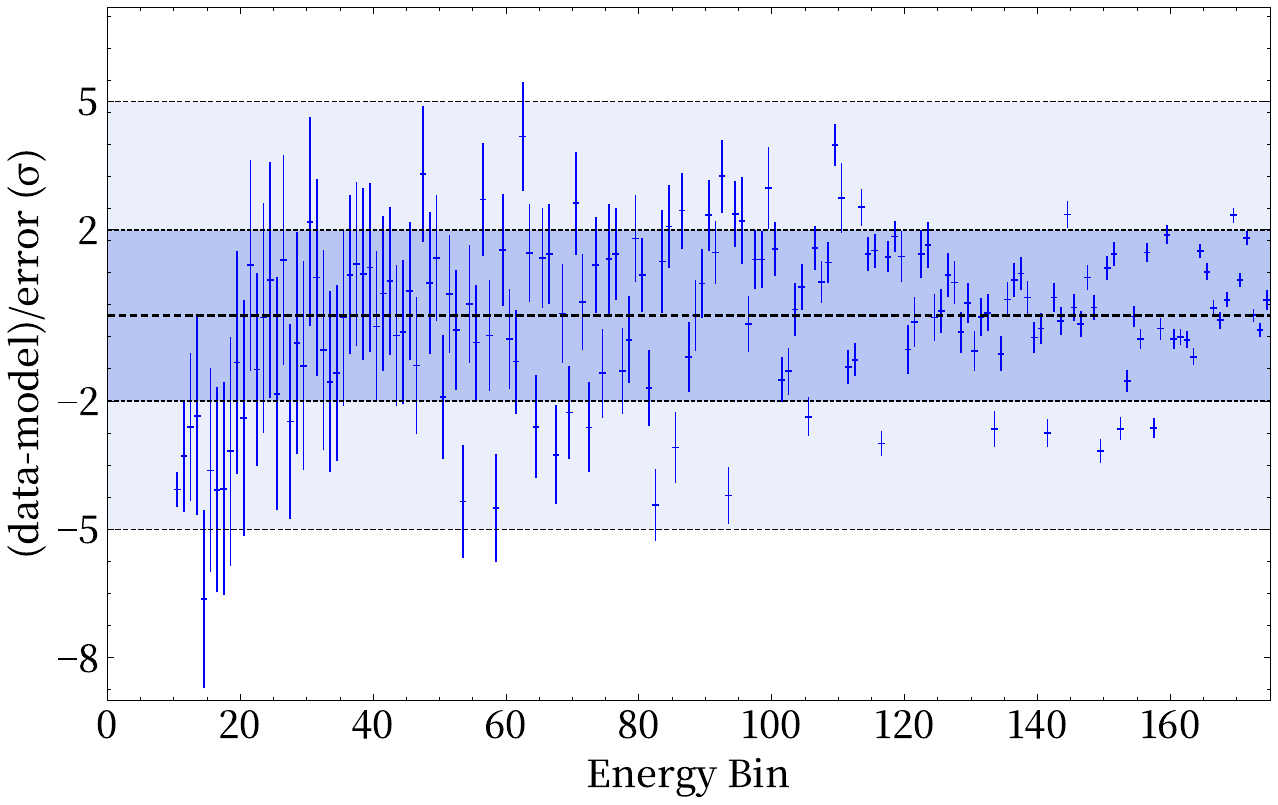}
  \caption{Distribution of Maximum Standardized Residual Values (MRV) across energy channels for 27 spectra. The distribution shows, for each energy channel (x-axis), the maximum absolute standardized residual found among the 27 cross-validation spectra. This distribution is used to assess the conservative upper bound of the systematic error of our correction method under different spectral configurations.} 
  \label{fig:11}    
\end{figure}

The primary source of systematic uncertainty in our framework is the residual dependence of the correction function $C(E)$ on the specific spectral shape used for its derivation (the reference spectrum $A$). To rigorously quantify this model-dependent systematic error, we analyze the results of the 27 cross-validation tests, where a fixed $C(E)$ was applied to a wide range of mismatched spectra ($B$).

We employ a suite of four metrics to provide a comprehensive assessment, with full results compiled in Table~\ref{tab:2}. These include two descriptive metrics—Cosine Similarity, which quantifies shape fidelity, and Euclidean Distance, which measures differences in both shape and amplitude—and two non-parametric goodness-of-fit tests: the Kolmogorov-Smirnov (K-S) and the more tail-sensitive Anderson-Darling (A-D) tests. The latter two provide a formal assessment of statistical significance.

The results in Table~\ref{tab:2} reveal a clear trend: the accuracy of the reconstruction degrades as the parameters of the target spectrum $B$ diverge from those of the reference spectrum $A$. This relationship is visualized in Figure~\ref{fig:10}, which shows that large differences in the high-energy power-law index $\beta$ and the characteristic energy $E_c$ are the most critical drivers of this performance degradation. As the mismatch in ($\beta, E_c$) increases, we observe a concurrent decrease in cosine similarity, an increase in Euclidean distance, and a general decline in the p-values from the statistical tests.

The statistical tests offer a more nuanced insight. The K-S test passes in 23 of 27 cases (85\%), confirming that the overall spectral shape is robustly recovered even under mismatched conditions. However, the more stringent A-D test, with a pass rate of 56\% (15 of 27 cases), indicates that minor but statistically significant deviations can emerge in the spectral tails for the most discrepant cases. This provides an honest quantification of the method's limitations.

Finally, to establish a conservative, energy-dependent upper bound for these systematic errors, we define the Maximum Residual Value (MRV) for each energy bin, $E_i$. The MRV is determined by identifying the single largest absolute residual found for that bin across all $j=1, \dots, 27$ cross-validation pairs ($B_j, B''_j$), as computed by:
\begin{equation}
\label{eq:mrv}
\text{MRV}(E_i) = \max_{j} \left| \frac{B_j(E_i) - B''_j(E_i)}{\sigma_j(E_i)} \right|,
\end{equation}
where the uncertainty $\sigma_j(E_i)$ is the statistical error $\sqrt{B_j(E_i)}$ \citep{draper1998applied, bevington2003data}. The resulting distribution of these 175 MRVs is plotted in Figure~\ref{fig:11}.The resulting distribution of these 175 MRVs is plotted in Figure~\ref{fig:11}. This distribution serves as our worst-case estimate for the systematic uncertainty as a function of energy. We find that 98\% of the MRV values are less than 5, meaning the maximum systematic deviation is contained within $\pm 5\sigma$ at this confidence level. This provides a robust and conservative cap on the systematic errors of our method.

\renewcommand{\arraystretch}{1.15} 
\setlength{\tabcolsep}{13pt} 
\begin{table*}[htbp] 
\centering 
\caption{Cross-Validation Metrics Comparing Reconstructed and Incident Spectra} 
\label{tab:2} 
\begin{tabular}{ccccccc} 
\hline 
\multicolumn{1}{c}{$\alpha$} & \multicolumn{1}{c}{$\beta$} & \multicolumn{1}{c}{$E_c$ (keV)} & \multicolumn{1}{c}{Cosine Similarity} & \multicolumn{1}{c}{Euclidean Distance} & \multicolumn{1}{c}{K-S p-value} & \multicolumn{1}{c}{A-D p-value}\\ 
\multicolumn{1}{c}{(1)} & \multicolumn{1}{c}{(2)} & \multicolumn{1}{c}{(3)} & \multicolumn{1}{c}{(4)} & \multicolumn{1}{c}{(5)} & \multicolumn{1}{c}{(6)} & \multicolumn{1}{c}{(7)}\\ 
\hline 
-0.66 & -1.90 & 640  & 0.9609 & 0.5159 & 0.3076 & 0.2041 \\
-0.66 & -1.90 & 1503 & 0.9600 & 0.2884 & 0.2993 & 0.1991 \\
-0.66 & -1.90 & 2250 & 0.9573 & 0.3455 & 0.0935 & 0.0239 \\
-0.66 & -2.34 & 640  & 0.9984 & 0.2127 & 0.3991 & 0.2121 \\
-0.66 & -2.34 & 1503 & 0.9983 & 0.2351 & 0.4270 & 0.2268 \\
-0.66 & -2.34 & 2250 & 0.9946 & 0.5189 & 0.3820 & 0.1393 \\
-0.66 & -2.85 & 640  & 0.9710 & 0.1476 & 0.0455 & 0.0342 \\
-0.66 & -2.85 & 1503 & 0.9789 & 0.3822 & 0.0590 & 0.0446 \\
-0.66 & -2.85 & 2250 & 0.9862 & 1.0642 & 0.0308 & 0.0152 \\
-1.17 & -1.90 & 640  & 0.9602 & 0.2961 & 0.3348 & 0.0584 \\
-1.17 & -1.90 & 1503 & 0.9596 & 0.0476 & 0.3988 & 0.0215 \\
-1.17 & -1.90 & 2250 & 0.9604 & 0.3353 & 0.2202 & 0.1810 \\
-1.17 & -2.34 & 640  & 0.9984 & 0.2263 & 0.4266 & {$>0.25$} \\
\textbf{-1.17} & \textbf{-2.34} & \textbf{1503} & \textbf{0.9992} & \textbf{0.0394} & \textbf{0.4291} & \bm{$>0.25$} \\ 
-1.17 & -2.34 & 2250 & 0.9985 & 0.2145 & 0.1582 & 0.0188 \\
-1.17 & -2.85 & 640  & 0.9696 & 0.1610 & 0.0434 & 0.0124 \\
-1.17 & -2.85 & 1503 & 0.9727 & 0.0415 & 0.0834 & 0.0234 \\
-1.17 & -2.85 & 2250 & 0.9818 & 0.4066 & 0.0342 & 0.0105 \\
-1.50 & -1.90 & 640  & 0.9593 & 0.3280 & 0.0847 & 0.0251 \\
-1.50 & -1.90 & 1503 & 0.9610 & 0.2938 & 0.2186 & 0.0660 \\
-1.50 & -1.90 & 2250 & 0.9599 & 0.2911 & 0.0532 & 0.0251 \\
-1.50 & -2.34 & 640  & 0.9984 & 0.2389 & 0.1496 & 0.0421 \\
-1.50 & -2.34 & 1503 & 0.9983 & 0.2268 & 0.3247 & 0.1841 \\
-1.50 & -2.34 & 2250 & 0.9985 & 0.1970 & 0.1606 & 0.0524 \\
-1.50 & -2.85 & 640  & 0.9706 & 0.1523 & 0.1003 & 0.0514 \\
-1.50 & -2.85 & 1503 & 0.9706 & 0.1808 & 0.0737 & 0.0402 \\
-1.50 & -2.85 & 2250 & 0.9751 & 0.1917 & 0.0613 & 0.0315 \\
\hline 
\end{tabular}

\begin{minipage}{\linewidth}  
\textbf{\textit{Note.}} Columns are: (1--3) Input Band function parameters for the incident spectrum. (4) Cosine similarity and (5) Euclidean distance between the incident and reconstructed spectra. A higher similarity and lower distance indicate better agreement. (6) p-value from the two-sample Kolmogorov-Smirnov (K-S) test. (7) p-value from the two-sample Anderson-Darling (A-D) test. The entire analysis utilizes a single correction function, $C(E)$, which was derived from the incident spectrum defined in the bolded row. Therefore, the bolded row represents the ideal self-consistency test. All other rows are cross-validation tests, where this fixed $C(E)$ is applied to spectra with mismatched parameters to test the method's generalizability. The p-value is reported as a lower bound as the test statistic falls outside the range for precise calculation.
\end{minipage}
\end{table*}

\section{Application and Validation Using GRB 221009A Data} \label{sec:5}
To validate our Monte Carlo  simulation-based data acquisition correction method, we apply it to the brightest gamma-ray burst GRB 221009A \citep{2023arXiv230301203A, 2024SCPMA..6789511Z}, observed by HEPP-H. We compare the corrected spectral results with independent measurements from GECAM-C over their shared energy range (2--5 MeV). This comparison, using observational data, serves to test the practical application of our correction method. 

\begin{figure}[htbp]
    \centering
    \begin{subfigure}{0.44\textwidth} 
        \includegraphics[width=\linewidth]{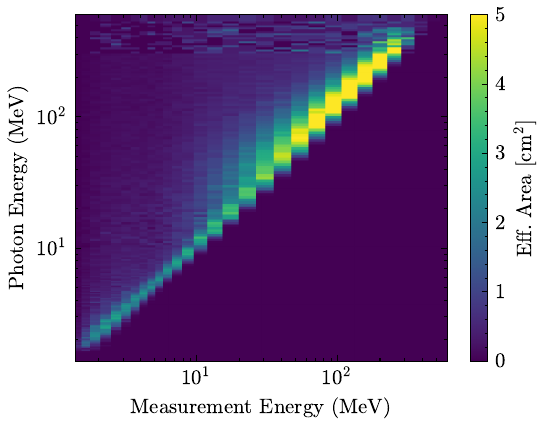}
        \caption{Response Matrix} 
        \label{fig:subfig1} 
    \end{subfigure}
    \hfill 
    \begin{subfigure}{0.42\textwidth} 
        \includegraphics[width=\linewidth]{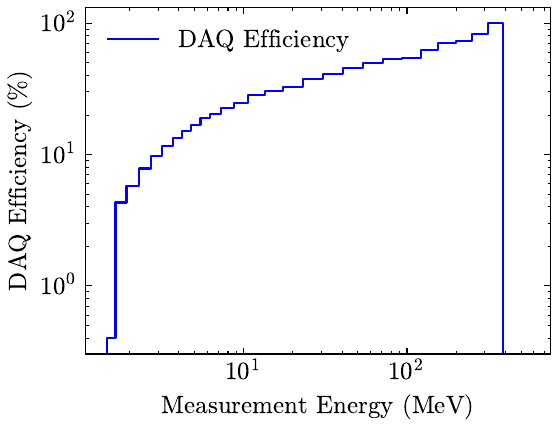} 
        \caption{correction Function} 
        \label{fig:subfig2} 
    \end{subfigure}

    \caption{\textbf{{(a):}} Instrument response matrix for HEPP-H, combining the RMF and ARF, representing the detector's modeled photon response. \textbf{{(b):}} Energy-dependent correction function $C(E)$ for HEPP-H, which accounts for the instrumental effects discussed in Section~\ref{sec:2}.
    } 
    \label{fig:12} 
\end{figure}

We analyze data from HEPP-H and GECAM-C during the 232--233 s interval of GRB 221009A's outburst. This high count rate event provides an ideal testbed for our data acquisition correction method. The GECAM-C detector, part of the GECAM series, offers a wide energy range, large field of view, and moderate localization capability \citep{2023NIMPA105668586Z, 2024NIMPA105969009Z}, making its observations a suitable independent validation reference for the HEPP-H inversion results. For robust comparison, we used low-gain data (0.7-5.5 MeV) from GECAM-C and restricted both HEPP-H and GECAM-C spectral data to the 2--5 MeV common energy interval. We began by loading the RMFs and ARFs for both detectors. Subsequently, the HEPP-H RMF was corrected using the correction function $C(E)$ shown in Figure~\ref{fig:12}, yielding the corrected RMF, RMF$'$, for subsequent spectral fitting.

To validate the efficacy of our data acquisition correction method, we performed comparative analysis and spectral inversion using observations of GRB 221009A from HEPP-H and GECAM-C.

First, using the Band function \citep{1993ApJ...413..281B}, we performed the spectral fitting of both detectors' data. For HEPP-H, we conducted comparative analysis using both the original (RMF) and corrected (RMF$'$) response matrices while keeping the ancillary response file (ARF) unchanged (Figure~\ref{fig:12}). The GECAM-C analysis employed its standard response matrices.

\begin{figure*}[ht!]  
    \centering
    \begin{subfigure}{0.48\textwidth}
        \includegraphics[width=\linewidth]{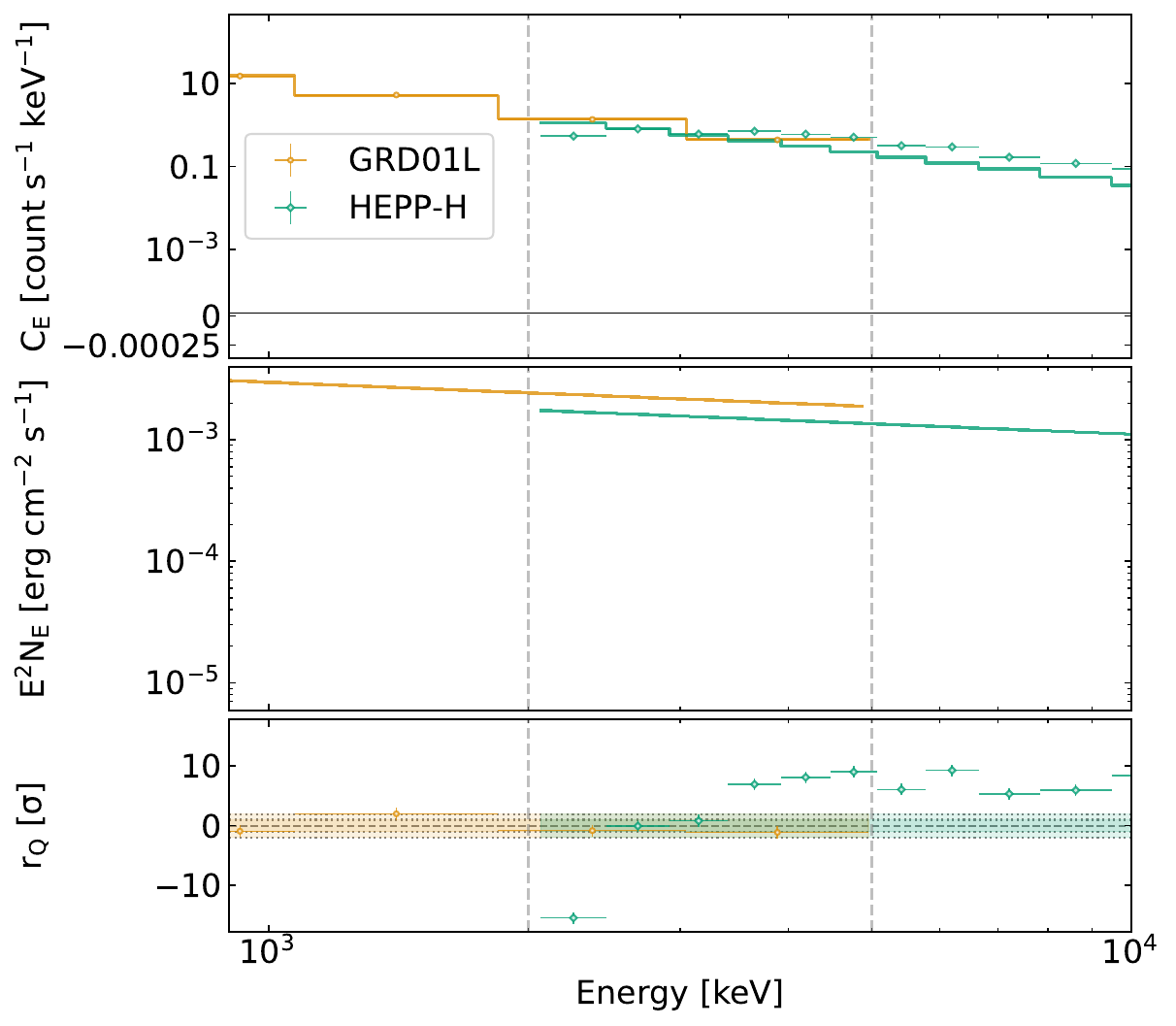}
        \label{fig:Peak search}
    \end{subfigure}\hfill
    \begin{subfigure}{0.48\textwidth}
        \includegraphics[width=\linewidth]{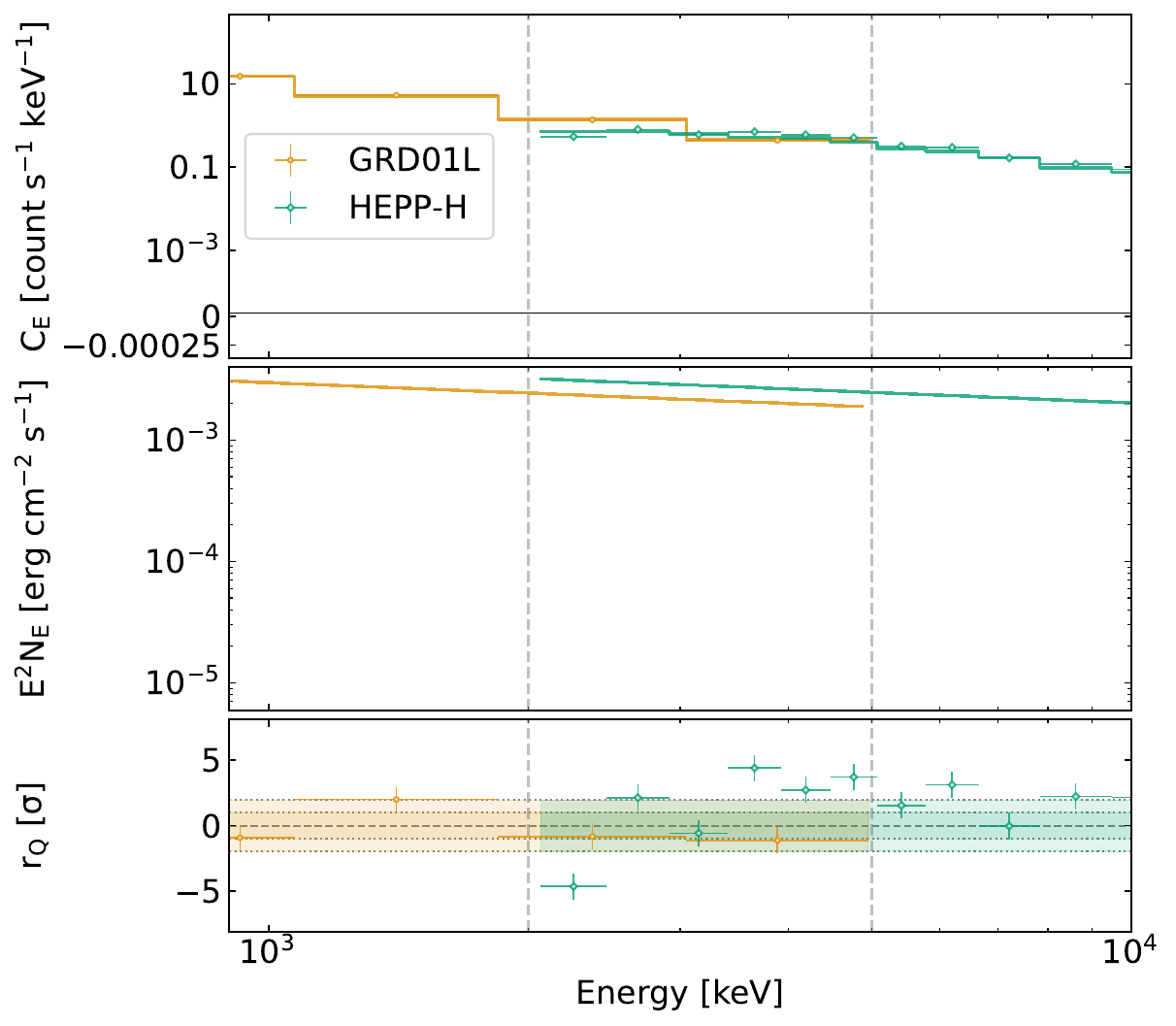}
        \label{fig:Dead time}
    \end{subfigure}
    \caption{ 
    Spectral fitting results for GRB\,221009A observed by HEPP-H and GECAM-C. \textbf{\textit{Left Panel:}} Fits using standard response matrices (RMF and ARF) for HEPP-H. \textbf{\textit{Right Panel:}} Fits using corrected response matrices (RMF$'$ and ARF) for HEPP-H. In both cases, GECAM-C data are fitted using its native response. The corrected HEPP-H response demonstrates improved spectral fitting with normalized residuals within $\pm5\sigma$ and yields consistent Band function parameters ($\alpha = -0.59 \pm 0.02$, $\beta = -2.29 \pm 0.10$, $E_{\rm c} = 392.75 \pm 12$\,keV). The gray dashed lines indicate the 2--5\,MeV energy range.
    }
    \label{fig:13}
\end{figure*}

After using the corrected RMF$'$(which incorporates $C(E)$), HEPP-H's spectral residuals improved significantly in the 2-5 MeV energy range (Figure~\ref{fig:13}). The corrected HEPP-H and GECAM-C observations are both well-described by the Band function model, demonstrating strong consistency within the common energy interval. The Band model parameters, including \textit{low-energy} power-law index $\alpha \approx -0.59$ and \textit{high-energy} power-law index $\beta \approx -2.29$, and characteristic energy $E_{c} \approx 392.75$ keV, were in agreement within their uncertainties. Normalized residuals fell within $\pm 5\sigma$, suggesting that the Band model provides an adequate representation of the observed data. The spectral fitting robustly demonstrates that our correction restores HEPP-H's spectral fidelity, providing a reliable benchmark for subsequent inversion analysis and cross-validation. It is crucial to underscore that our method, while applying a correction function to the observations, fundamentally remains an iterative spectral fitting technique. The correction parameters, derived from MC simulations, are integrally incorporated within this iterative process to enhance the spectral fidelity and reliability of the resulting analysis.

Second, we validated our correction and deconvolution methods by inverting the HEPP-H spectrum of GRB 221009A. The inversion comprised two key steps: 
\begin{itemize}
    \item Applying the correction function $C(E)$ (Section~\ref{subsubsec:C(E)}) to mitigate instrumental effects; 
    \item A machine learning-based deconvolution (Section~\ref{subsubsec:CNN_Inversion}) aimed at accounting for model dependence problem.
\end{itemize}

\begin{figure}[h] 
  \centering 
  \includegraphics[width=0.47\textwidth]{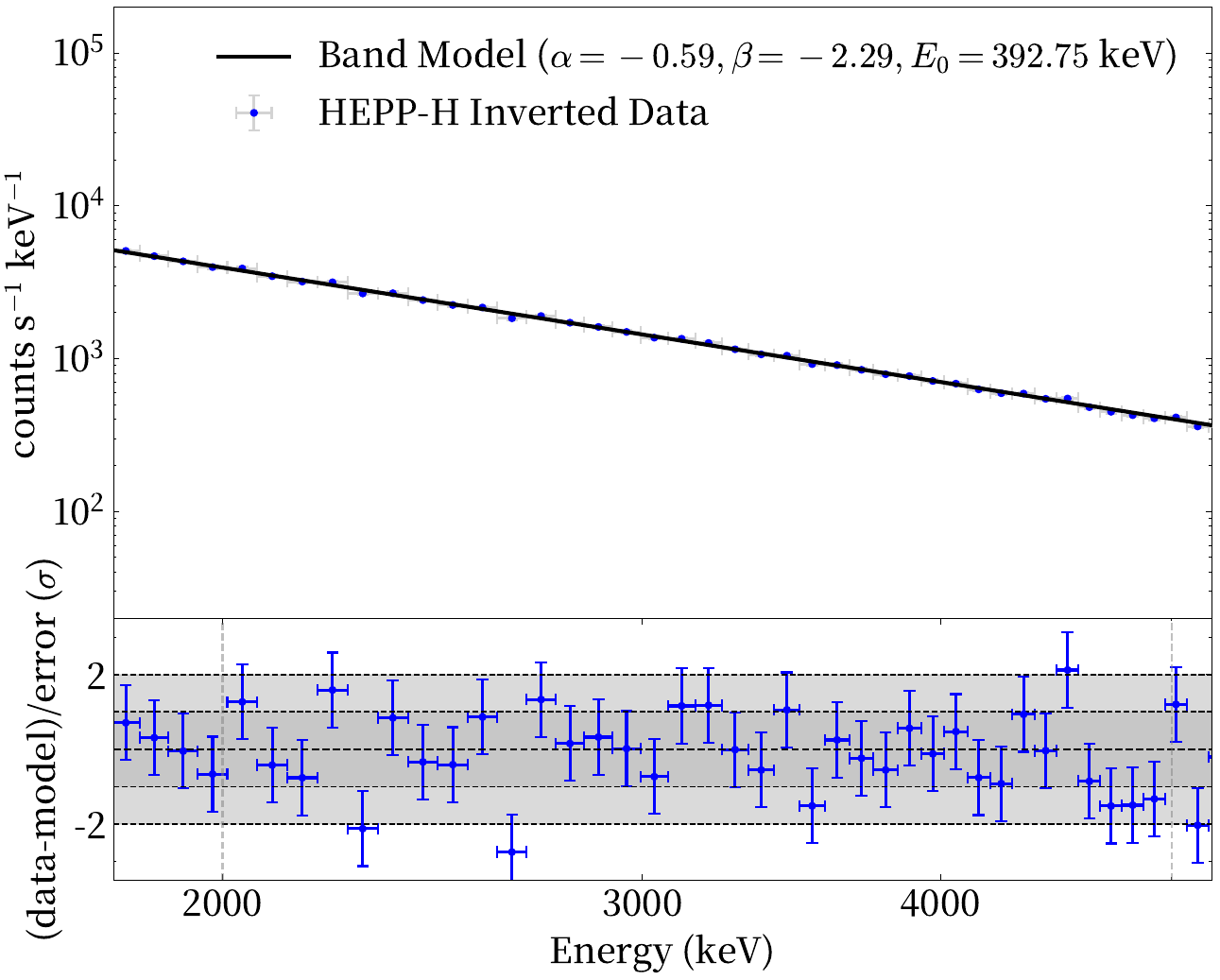}
  \caption{The inverted HEPP-H energy spectrum (gray points) compared with a best-fit Band function model (black solid line; $\alpha = -0.59$, $\beta = -2.29$, $E_0 = 392.75$\,keV). The data points are shown with their statistical uncertainties, propagated through the entire analysis pipeline. The model parameters were derived from a separate forward-fitting analysis. The lower panel displays the corresponding normalized residuals, calculated as (data$-$model)/error. These residuals are symmetrically distributed around zero and are predominantly confined within the $\pm$2$\sigma$ region, indicating excellent statistical agreement between our model-independent inversion and the traditional model-fitting result.} 
  \label{fig:14}    
\end{figure}

We quantitatively assessed our combined approach by comparing the HEPP-H inverted spectrum with the spectral fitting Band model ($\alpha$ = -0.59, $\beta$ = -2.29, $E_{c}$ = 392.75 keV). Figure~\ref{fig:14} shows that across the 2--5 MeV range, the normalized residuals between the inverted spectrum and the Band model are confined to $\pm 2\sigma$. This is a significant improvement compared to the $\pm 5\sigma$ residuals observed when only applying the $C(E)$ correction (i.e., in the spectral fitting stage using RMF$'$ without subsequent deconvolution). This reduction in residual magnitude demonstrates that our methods recover the intrinsic spectrum while remaining consistent with standard analysis techniques.

\section{Conclusions and Discussion} \label{sec:6}
\subsection{Conclusion} \label{subsec:Conclusions}
This study presents a novel correction framework for mitigating instrumental distortions in the HEPP-H detector during high-count-rate GRB observations. Through comprehensive Monte Carlo simulations and observational validation, we demonstrate significant improvements in spectral recovery accuracy with four principal findings:

\begin{enumerate}
    \item \textbf{Simulation Validation:} Under idealized conditions, our correction method recovers simulated spectra with high fidelity. Normalized residuals remain within expected statistical fluctuations, with 90.9\% of residuals falling within $\pm 2\sigma$ (Figure~\ref{fig:9}), demonstrating the method's accuracy in controlled conditions.
    
    \item \textbf{Cross-Validation and Generalization:}  Cross-validation indicates that the correction factor C(E) exhibits good generalization across a range of GRB spectral parameters relevant to the brightest GRB. It effectively handles cases where spectra belong to the same spectral class but have different parameters, showing a robust correction effect across the 27 sets of spectral parameters tested. Rigorous statistical tests confirm this robustness: the Kolmogorov-Smirnov test indicates no significant deviation for 23 of the 27 cases (85\%), and even the more stringent Anderson-Darling test passes for a majority of the configurations (15 of 27 cases, 56\%). A detailed breakdown of these metrics is provided in Table~\ref{tab:2}.

    \item \textbf{CNN Deconvolution Performance:} Experimental results show that the CNN-based deconvolution method introduced in this paper exhibits numerical stability and nonlinear mapping capability, offering advantages in spectral inversion efficiency compared to the traditional Tikhonov regularization algorithm (see Figure~\ref{fig:6} and Figure~\ref{fig:8}). A key feature of our framework is the construction of an explicit inverse energy response matrix—extending over a broader energy range than the analysis band (see Figure~\ref{fig:7} and Figure~\ref{fig:8})—which is derived statistically from simulations. This statistically-derived matrix effectively embodies the linear component of the more complex, nonlinear inverse transformation learned by the CNN, as both are fundamentally rooted in extracting statistical relationships from the same underlying simulation data.

    \item \textbf{Observational Validation:} To evaluate the efficacy of the correction method on observational data, we performed forward fitting and spectral inversion on HEPP-H observations of GRB\,221009A in the 2--5 MeV range (Figure~\ref{fig:13}, Figure~\ref{fig:14}). The corrected energy spectrum shows agreement with independent GECAM-C measurements in the same energy band, and the systematic error of the correction method is $\pm 5\sigma$ (Figure~\ref{fig:11}). The combined correction and deconvolution reduces instrumental effects in HEPP-H data at high count rates, validating the method for HEPP-H applications (Figure~\ref{fig:14}).

\end{enumerate}

\subsection{Discussion} \label{subsec:Discussions}
The data correction and deconvolution method presented in this study has been validated through simulations and observational data. The method reduces instrumental effects in HEPP-H data collected at high count rates, improving the accuracy of reconstructed energy spectra. Compared to traditional approaches, the convolutional neural network-based deconvolution shows improved numerical stability and inversion efficiency for GRB spectral analysis.

As discussed in Section~\ref{subsec:S E}, the systematic error assessment based on maximum residual values provides an upper limit estimate but may overestimate actual errors as it does not utilize the full residual distribution. Future work will implement goodness-of-fit tests, including Chi-Squared Test ($\chi^2$; \citet{Pearson1992}) and Likelihood Ratio Test \citep{Anderson01121954}, to better quantify both the correction method's performance and systematic uncertainties.

This study employs the quantile loss function ($\tau = 0.5$), equivalent to Least Absolute Deviation (LAD) regression. As shown in \citet{10.1007/978-981-15-0947-6_34} and our residual analysis (Figure~\ref{fig:9}), LAD regression reduces the influence of outliers compared to Ordinary Least Squares (OLS). For energy spectrum inversion, where $\gamma$-ray events and electronic noise introduce outliers in detector data, the quantile loss function improves inversion robustness against these artifacts.

The generalization capabilities of our framework warrant careful discussion, as the cross-validation tests reveal that the method's performance is not entirely independent of the reference spectrum used to derive the correction function $C(E)$. While the method proves highly effective for spectra within the same class as the reference, its accuracy degrades as the spectral shape diverges. This dependence is quantitatively captured by the comprehensive metrics in Table~\ref{tab:2} and visualized in the parameter space of Figure~\ref{fig:10}. Specifically, as the \textit{high-energy} power-law index $\beta$ and characteristic energy $E_c$ of a target spectrum deviate significantly from the reference model, we observe a concurrent decrease in cosine similarity, an increase in Euclidean distance, and a decline in the goodness-of-fit p-values.

Our statistical analysis provides a nuanced view of this effect. The Kolmogorov-Smirnov (K-S) test results confirm that the framework is robust in recovering the overall spectral shape, with the vast majority of cases showing no statistically significant deviations. However, the more stringent Anderson-Darling (A-D) test, with its sensitivity to the spectral tails, fails for the most discrepant cases. This implies that for GRBs with spectra substantially different from the basis upon which $C(E)$ was constructed, the accuracy of the correction itself may degrade, leading to minor but statistically significant systematic residuals, particularly in the spectral tails. Addressing this residual model dependence is therefore a key direction for future work.

Future research to enhance the accuracy and general applicability of this correction method can focus on:
\begin{itemize}
    \item Refining Monte Carlo simulations to more accurately model detector response characteristics, providing a foundation for improved spectral correction.
    \item Automating simulation-fitting iterations to eliminate manual intervention, achieve more exhaustive convergence, and reduce dependence on initial assumptions.
    \item Developing more adaptive and robust deconvolution algorithms, potentially exploring advanced deep learning or adaptive methods.
    \item Developing deconvolution methods aimed at eliminating or factoring out the influence of assumed incident spectral shapes inherent in the response characterization.
    \item Exploring more comprehensive methods for quantifying systematic errors to better assess and control uncertainties and improve result reliability.
    \item Constructing more adaptive correction functions, for instance by dynamically adjusting $C(E)$ based on GRB spectral shape, and broadening the range of astrophysical source spectral models incorporated into future simulations to enhance the generalizability of $C(E)$.
\end{itemize}

In summary, while the framework presented in this work provides a practical solution for effectively correcting HEPP-H data during high-count-rate GRB observations, its fundamental strength lies in the universal applicability of its core components. The data acquisition correction method of Section~\ref{sec:2} is broadly applicable to various X-ray, gamma-ray, and particle detectors facing high-rate challenges. Its essence lies in a simulation-based iterative spectral fitting process that explicitly incorporates \textbf{instrumental correction parameters}; this makes it particularly powerful when instrumental effects significantly distort the data, for instance, when the dead time fraction is significant. Furthermore, the inverse deconvolution method presented in Section~\ref{sec:3} represents a versatile and numerically stable approach to spectral unfolding. Currently, its strength lies in effectively inverting spectra that belong to a known spectral class even with significant variations in their specific model parameters. However, the accuracy of this data-driven method is contingent not only on the precision of the instrument's energy response matrix characterization, but also on the similarity between the incident spectral shape being analyzed and the spectral forms used to derive the inverse operator. A key goal for future development is therefore to enhance the method's model independence, aiming to decouple the deconvolution process from the specific spectral assumptions inherent in the derivation of the inverse response.

Together, these methods provide a robust and adaptable toolkit for spectral inversion and correction applicable across many current and future high-energy astrophysics missions encountering complex instrumental effects. By enabling more accurate spectral analysis, particularly for high-energy transients, this work allows for the full exploitation of the scientific potential of X-ray, gamma-ray, and particle detectors, significantly contributing to GRB astrophysics and providing essential technical guidance for future observational campaigns.


\begin{acknowledgments}
This work was supported by grants from the National Natural Science Foundation of China (NNSFC) (12173038, 42474222, 1177525) and the Strategic Priority Research Program of the Chinese Academy of Sciences (grant XDA15360000, Nos. XDA15360102 and E02212A02S). We acknowledge the data support for this research provided by the China Centre for Resources Satellite Data and Application and the National Institute of Natural Hazards. 
\end{acknowledgments}

\end{document}